\documentclass[letterpaper,journal]{IEEEtran}
\usepackage{amsmath,amsfonts}
\usepackage{algorithmic}
\usepackage{algorithm}
\usepackage{array}
\usepackage[caption=false,font=normalsize,labelfont=sf,textfont=sf]{subfig}
\usepackage{textcomp}
\usepackage{stfloats}
\usepackage{url}
\usepackage{verbatim}
\usepackage{graphicx}
\graphicspath{{./Fig/}}
\usepackage{epstopdf}
\usepackage{cite}
\usepackage{bm}
\usepackage{float}
\usepackage{amssymb}
\usepackage{adjustbox} 
\usepackage{caption}
\ifdefined\pdfoutput
	\pdfoutput=1
\fi
\allowdisplaybreaks[4]

\newcounter{theorem}
\newcounter{lemma}
\newcounter{remark}
\newenvironment{theorem}{\refstepcounter{theorem}\par\noindent\itshape\textbf{Theorem \thetheorem:} \upshape}{\par}
\newenvironment{lemma}{\refstepcounter{lemma}\par\noindent\itshape\textbf{Lemma \thelemma:} \upshape}{\par}
\newenvironment{remark}{\refstepcounter{remark}\par\noindent\itshape\textbf{Remark \theremark:} \upshape}{\par}
\newenvironment{proof}{\par\noindent\itshape\textbf{Proof:} \upshape}{\hfill$\blacksquare$\par}

\begin{document}
	%
	\title{A Framework for Uplink ISAC Receiver Designs: Performance Analysis and Algorithm Development}


		
		\author{Zhiyuan Yu, Hong Ren, \IEEEmembership{Member, IEEE}, Cunhua Pan, \IEEEmembership{Senior Member, IEEE}, Gui Zhou, \IEEEmembership{Member, IEEE}, Dongming Wang, \IEEEmembership{Member, IEEE},  Jiangzhou Wang, \IEEEmembership{Fellow, IEEE}, 
			\vspace{-0.2cm}
			\thanks{Part of this work has been accepted by ICC 2025\cite{11160919}. The work of Cunhua Pan was supported in part by the Scientific Research Innovation Capability Support Project for Young Faculty under Grants 3204002501C3 and U40 and in part by the National Natural Science Foundation of China under Grant U25A20393. The work of Hong Ren was supported in part by the National Natural Science Foundation of China under Grant No. 62471138. The work of Jiangzhou Wang was supported in part by the National Natural Science Foundation of China under Grant 62350710796. 
			
				Zhiyuan Yu, Hong Ren, Cunhua Pan, and Dongming Wang are with the National Mobile Communications Research Laboratory, Southeast University, Nanjing, China (e-mail: \{zyyu, hren, cpan, wangdm\}@seu.edu.cn). Jiangzhou Wang is with the National Mobile Communications Research Laboratory, Southeast University, Nanjing, China, and also with Purple Mountain Laboratories, Nanjing, Jiangsu 211119, China (e-mail: j.z.wang@seu.edu.cn). Gui Zhou is with the School of Electronic Information and Communication, Huazhong University of Science and Technology (HUST), Wuhan, China (e-mail: gui\_zhou@hust.edu.cn). 
				
				\emph{Corresponding authors: Hong Ren and Cunhua Pan.}
				
			}	
		}
		\maketitle


		
		%


		\maketitle
		\vspace{-1cm}

		\begin{abstract}			
			Uplink integrated sensing and communication (ISAC) systems have recently emerged as a promising research direction, enabling simultaneous uplink signal detection and target sensing. {In this paper, we propose the flexible projection (FP)-type receiver that unifies the projection-type receiver and the successive interference cancellation (SIC)-type receiver by using a flexible tradeoff factor to adapt to dynamically changing uplink ISAC scenarios.} The FP-type receiver addresses the joint signal detection and target response estimation problem through two coordinated phases: 1) Communication signal detection using a reconstructed signal whose composition is controlled by the tradeoff factor, followed by 2) Target response estimation performed through subtraction of the detected communication signal from the received signal. With adjustable tradeoff factors, the FP-type receiver can balance the enhancement of the signal-to-interference-plus-noise ratio (SINR) with the reduction of correlation in the reconstructed signal for communication signal detection. The pairwise error probability (PEP) expressions are analyzed for both the maximum likelihood (ML) and the zero-forcing (ZF) detectors, revealing that the optimal tradeoff factor should be determined based on the adopted detection algorithm and the relative power of the sensing and communication (S\&C) signals. A homotopy optimization framework is first applied for the FP-type receiver with a fixed tradeoff factor. This framework is then extended to develop the dynamic flexible projection (DFP)-type receiver, which iteratively adjusts the tradeoff factor for improved algorithm performance and environmental adaptability. Finally, we show that the length of the jointly processed signal should scale with the antenna size to fully unleash the potential of the uplink ISAC receiver.

		\end{abstract}
		
		

		\IEEEpeerreviewmaketitle
		
		\section{Introduction}

		Low-altitude economy (LAE) is expected to support a variety of applications in transportation, environmental monitoring, agriculture, and entertainment, thereby generating significant economic and social value. ISAC \cite{10418473,9171304}, a key technology for sixth-generation (6G) wireless networks, has emerged as an effective solution for supporting LAE \cite{156489, 156489566}. The continuous development of ISAC enables base stations (BSs) to communicate with designated unmanned aerial vehicles (UAVs) while simultaneously sensing the location and velocity of other aerial targets, thereby establishing dual-functional capabilities for heterogeneous service demands. This dual-functional integration offers particular advantages in enhancing operational efficiency and safety assurance for latency-sensitive applications such as real-time logistics tracking and aerial surveillance systems.
		
		In contrast to traditional terrestrial networks that primarily emphasize downlink services, UAVs require sustained uplink transmission for multi-modal sensing data, such as video streams and environmental sampling data, leading to higher uplink data-rate demands. The coexistence of these uplink-dominated communication services with continuous target-sensing requirements presents substantial challenges for future ISAC networks. Although these S\&C services can be supported by allocating different subcarrier or time-resource blocks, doing so leads to low resource utilization. To better exploit spectrum resources for enhanced S\&C performance, particularly for continuous sensing, the uplink ISAC receiver is expected to detect the uplink communication signal while simultaneously estimating target-related parameters from the echoes. However, these concurrent S\&C tasks introduce significant mutual interference, making receiver design for uplink ISAC systems both intriguing and challenging.


		Several initial investigations have been conducted on uplink ISAC systems, focusing on system architecture design \cite{8999605,9618653}, performance analysis \cite{SICOuyang, NOMA_ISAC2, 10608079, 10542219}, and transceiver beamforming \cite{10472418, 10158711, 10557567}. These studies on architecture design explored the potential of uplink ISAC systems and provided initial attempts to address the challenging joint signal detection and sensing estimation problem. Performance-analysis studies mainly investigated the tradeoff among outage probability, sensing rate, and their asymptotic behaviors. These studies indicated that, despite mutual interference, ISAC receivers can offer increased degrees of freedom (DoFs) for both S\&C functionalities compared to frequency-division S\&C systems. Meanwhile, several contributions focused on transceiver design in uplink ISAC, exploring the S\&C performance tradeoff and interference mitigation through power allocation and beamforming techniques. 
		
		The aforementioned research primarily focused on exploring the potential of uplink ISAC systems with the SIC-type receiver. Notably, in studies on architecture design \cite{8999605,9618653}, communication signal detection and radar target estimation were performed sequentially. Furthermore, performance metrics used in analyses and transceiver beamforming, such as sensing rate \cite{SICOuyang} and communication and sensing SINR \cite{10158711}, were derived by treating mutual interference as noise, which is a key characteristic of the SIC-type receiver. However, the SIC-type receiver represents a heuristic approach to the joint signal detection and sensing problem and has been shown to be suboptimal in simplified system setups \cite{Fan}. Furthermore, the SIC-type receiver requires a significant power difference between S\&C signals, making it unsuitable when the powers of the communication and sensing signals are comparable.
		
		To further harness the potential of uplink ISAC systems, it is crucial to design more efficient receivers to mitigate mutual interference or, equivalently, solve the joint uplink communication signal detection and target response estimation problem. Although the considered joint problem is related to problems such as joint channel estimation (CE) and signal detection \cite{9018199}, superimposed-pilot-based systems \cite{9508784}, and interference elimination in radar-communication coexistence (RCC) systems \cite{8233171,9420308}, their objectives and design principles differ fundamentally. Specifically, in joint CE and signal detection problems, a turbo-like structure was employed in which CE and signal detection were iteratively refined, using data to enhance CE accuracy. In superimposed-pilot-based systems, the transceiver is jointly optimized using a neural network to overcome the effect of imperfect channel state information (CSI). In RCC systems, the receiver was designed by exploiting frequency sparsity to suppress radar interference for communication signal detection \cite{8233171}, or vice versa \cite{9420308}.
		
		More relevant to this study, the joint communication signal detection and target estimation problem was investigated in \cite{10663294}, where the projection-type receiver was proposed for uplink ISAC systems. In \cite{10663294}, the joint communication signal detection and target estimation problem was transformed into an equivalent two-step problem through projection. Specifically, the communication signal was detected in a transformed signal detection problem, and sensing estimation was performed after the communication symbols were detected. Unlike the SIC-type receiver, which treats mutual interference as noise, the projection-type receiver fully eliminates sensing interference during communication signal detection, making it possible to achieve high S\&C performance simultaneously. However, the projection-type receiver still faces several implementation challenges, such as rank deficiency and high computational complexity introduced by joint estimation of the communication signal across multiple snapshots. Furthermore, two fundamental problems remain unaddressed in uplink ISAC receiver design: \textbf{(1)} Directly applying existing multiple-input multiple-output (MIMO) signal detection algorithms to the transformed communication signal detection problem leads to either high computational complexity or poor performance when rank deficiency occurs. How can the uplink ISAC receiver be designed with tailored low-complexity algorithms that provide performance guarantees? \textbf{(2)} The power of the S\&C signal directly influences the strength of mutual interference, necessitating different receiver types to achieve optimal performance. However, the received power of the sensing signal is determined by the environment and target movement, which may change rapidly and is not known to the receiver. In this case, is it possible to develop an environment-adaptive receiver that can operate effectively with varying S\&C power?
		
		Against the above background, our contributions are summarized as follows:

		\begin{enumerate}
			\item We consider two types of uplink ISAC systems that employ different communication channel acquisition protocols. We then investigate the joint signal detection and target response estimation problem in uplink ISAC systems and introduce a general framework termed the FP-type receiver with a tradeoff factor. In the proposed FP-type receiver, the communication signal is detected using a reconstructed signal formed by combining two signal components of the received signal: one from the complement space of the radar waveform and the other from the signal space of the radar waveform, where the tradeoff factor controls the ratio of the signal space component. The FP-type receiver then performs target response estimation by subtracting the estimated communication signal from the received signal.
			\item We show that decreasing the tradeoff factor increases the signal-to-interference-plus-noise ratio (SINR) of the signal detection problem, while also amplifying the undesired correlation between the received signals. Then, the pairwise error probabilities (PEPs) of the FP-type receiver are studied for the maximum likelihood (ML) detector and zero-forcing (ZF) detector, respectively. The overall performance loss of a specific detector compared to the communication-only systems is characterized by the computation loss (C-Loss) and the mutual interference loss (MI-Loss). 
			\item Building on a MIMO signal detection algorithm based on homotopy optimization, we propose the DFP-type receiver. The proposed algorithm transforms the formulated signal detection problem into a continuous optimization problem. By adjusting the tradeoff factor in each iteration, the DFP-type receiver provides a smoother approximation to the original joint signal detection and target estimation problem and offers superior environmental adaptability compared to the traditional FP-type receiver.
			\item Simulation results demonstrate the effectiveness of the proposed design. The angle estimation error and the bit error rate (BER) of the proposed DFP-type receiver are smaller than those of the projection-type receiver and the SIC-type receiver. Finally, we show that the length of the jointly processed signal should scale linearly with the number of antennas.
			
		\end{enumerate}

		The remainder of this paper is organized as follows. In Section II, we present the system model for uplink ISAC. Related work is reviewed in Section III. A general receiver design principle and the corresponding performance evaluation are presented in Section IV. A low-complexity algorithm is developed in Section V. Finally, Sections VI and VII present the numerical results and conclusions, respectively.
		
		\emph{Notations}: 
		${{\mathbb{C}}^{M \times N }}$ denotes the set of $M \times N$ complex matrices.  ${{\mathbb{E}}}{\left[\cdot\right]} $ denotes the expectation operation. ${\left\| {\mathbf{x}} \right\|_2}$ denotes the 2-norm of vector ${\mathbf{x}}$.  ${\mathbf{X}} \otimes {\mathbf{Y}}$ denotes the Kronecker product between $\mathbf X$ and $\mathbf Y$. $\operatorname{Rank} (\mathbf{X})$, ${\left\| {\mathbf{X}} \right\|_F}$, ${\left\| {\mathbf{X}} \right\|_2}$,  ${\rm{Tr}}\left( {\mathbf{X}} \right)$,  $\lambda_i(\mathbf{X})$, $\omega_j(\mathbf{X})$  denote the rank, Frobenius norm, spectral norm, trace, the $i$-th largest eigenvalue, and the $j$-th largest singular value of ${\mathbf{X}}$, respectively. $\operatorname{vec}(\mathbf{X})$ denotes the vectorization operation of the matrix $\mathbf{X}$ and $\operatorname{Unvec}(\mathbf{x})$ denotes its inverse operation.  $\mathbf{I}_N$ denotes the  $N \times N$ identity matrix.   ${\cal C}{\cal N}({\mathbf{0}},{\mathbf{I}})$ represents a circularly symmetric complex Gaussian random vector with zero mean and identity covariance matrix.  ${\left( \cdot \right)^{\rm{T}}}$, ${\left( \cdot \right)^{*}}$,  ${\left(  \cdot \right)^{\rm{H}}}$, $\left(  \cdot \right)^{-1}$, and $\left(  \cdot \right)^{\dagger}$ denote the transpose, conjugate, Hermitian, inverse, and pseudoinverse operators, respectively. The convex hull of a non-empty set $ \mathcal{X} $ is denoted by $ \text{conv}(\mathcal{X}) $.

			\vspace{-0.3cm}
			\section{System Model}
			We consider an uplink narrowband MIMO ISAC system shown in Fig. 1\footnote{We adopt this assumption for simplicity, and the proposed algorithm can be readily extended to multi-subcarrier scenarios.}, consisting of $K$ single-antenna communication users (CUs) and multiple targets. We assume that a transmitter equipped with $M_t$ transmit antennas is used to transmit the sensing signal, and a receiver equipped with $M_r$ receive antennas receives both the uplink communication signal and the radar echo, where $K\le M_r$. In Fig. 2, we present the frame structure of the considered uplink ISAC systems, which is divided into two phases. Specifically, Phase I is designed for uplink CE to acquire CSI through uplink pilots, while Phase II performs joint target sensing and uplink data detection over multiple snapshots. Depending on whether the uplink pilots (UPs) are overlapped with the sensing signal at the receiver, the uplink ISAC systems are classified into two types, i.e., Type I and Type II uplink ISAC, respectively. 
			
			\begin{figure*} 
				\centering
				\begin{minipage}[t]{0.49\textwidth} 
					\centering
					\includegraphics[width=8.5cm]{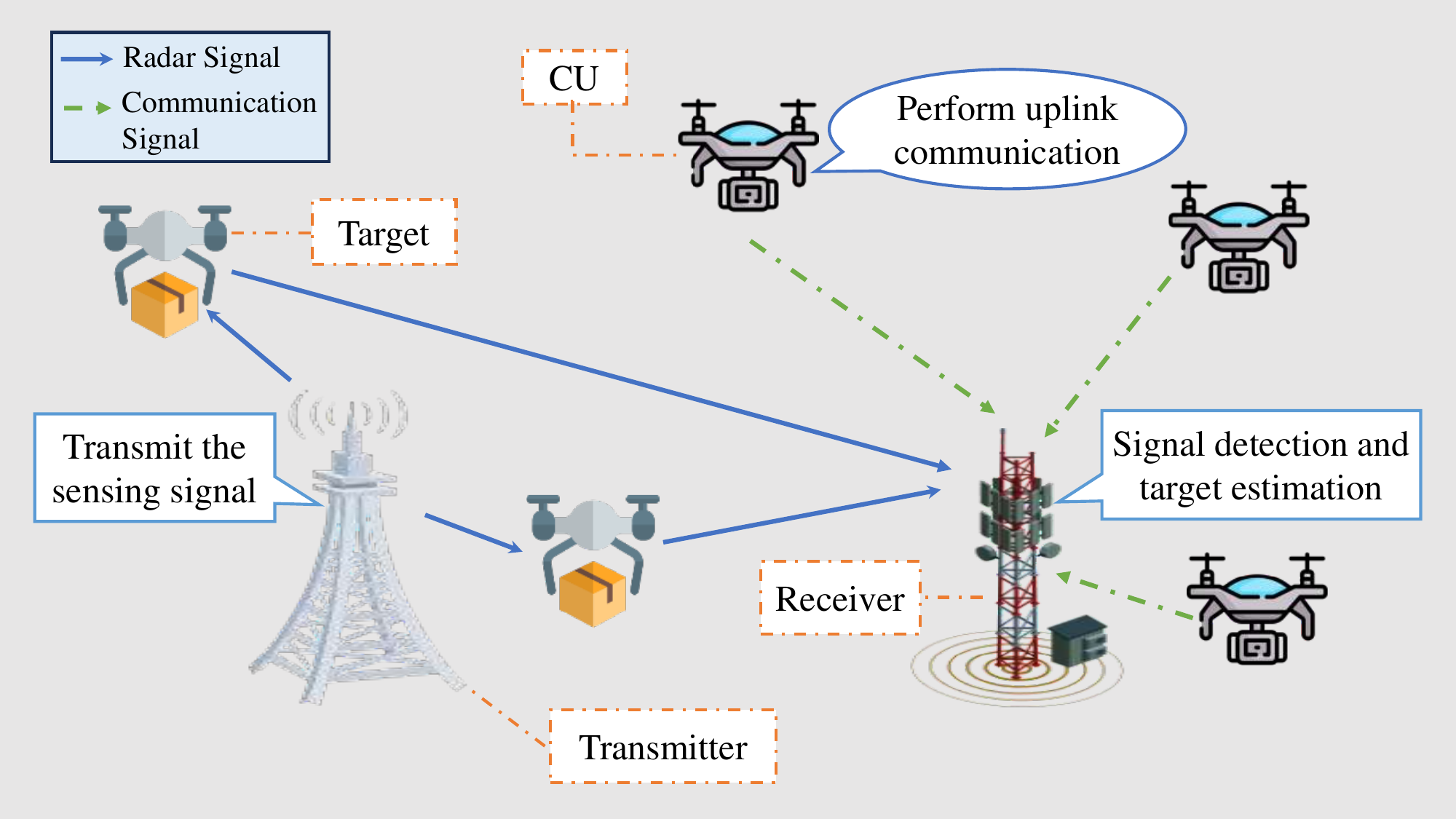}
					\caption{\small{Considered uplink ISAC systems}}
				\end{minipage}
				\begin{minipage}[t]{0.49\textwidth} 
					\centering
					\includegraphics[width=9cm]{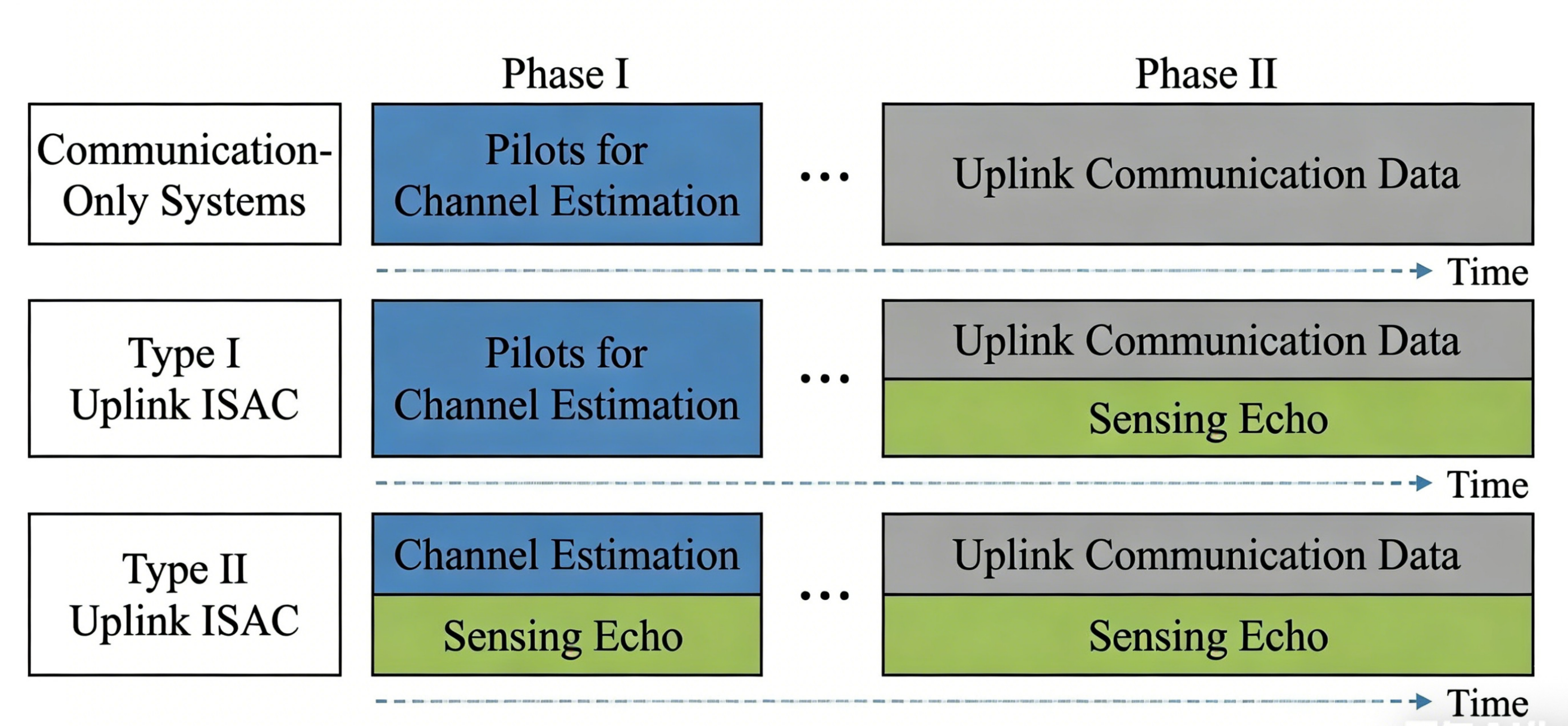}
					\caption{\small{Frame structure of the uplink ISAC systems}}
				\end{minipage}
				\vspace{-0.3cm}
			\end{figure*}
			

			\subsection{Phase I: Uplink Channel Estimation}

Acquiring accurate CSI is crucial for reliable communication. In the following, we briefly discuss two distinct CE designs and their respective advantages. In the Type I architecture, UPs are transmitted prior to the execution of sensing tasks, ensuring strict time-domain orthogonal transmission between the UPs and the sensing signals. Defining $L_p$ as the {length of the UPs}, the received pilot signal matrix $\mathbf{Y}_{\rm{p}} \in \mathbb{C}^{M_r \times L_p}$ can be modeled as
\begin{equation}
	\mathbf{Y}_{\rm{p}} = \mathbf{H}_{\rm{c}} \mathbf{X}_{\rm{p}} + \mathbf{N}_{\rm{p}},
\end{equation}
where $\mathbf{H}_{\rm{c}} \in \mathbb{C}^{M_r \times K}$, $\mathbf{X}_{\rm{p}} \in \mathbb{C}^{K \times L_p}$, and $\mathbf{N}_{\rm{p}} \in \mathbb{C}^{M_r \times L_p}$ denote the communication channel matrix between the CUs and the BS, the transmitted UP matrix, and the additive noise matrix at Phase I, respectively. After the signal is received, standard CE methods, such as least squares (LS) estimation \cite{4267831}, can be applied to obtain the estimated channel $\hat{\mathbf{H}}_{\rm{c}}$.

Conversely, the Type II architecture represents a relatively novel scheme for joint channel estimation and sensing by reusing the uplink pilot transmission time slots for sensing \cite{11020763}. In this case, the UPs are superimposed with the sensing signals, and the received signal matrix in Phase I is given by
\begin{equation}
	\mathbf{Y}_{\rm{p}} = \mathbf{H}_{\rm{c}} \mathbf{X}_{\rm{p}} + \mathbf{H}_{\rm{r}} \mathbf{X}_{\rm{r,p}}+ \mathbf{N}_{\rm{p}},
\end{equation}
where $\mathbf{X}_{\rm{r,p}} \in \mathbb{C}^{M_t \times L_p}$ represents the sensing signal matrix transmitted concurrently with the UPs, and $\mathbf{H}_{\rm{r}} \in \mathbb{C}^{M_r \times M_t}$ denotes the target response matrix, given by
\begin{equation}\label{chanel_los}
	{\mathbf{H}}_{ {\rm{r}}}=\sum_{p=1}^{P} b_{ p}  
	\mathbf{a}\left(M_{r}, \phi_{p}\right) \mathbf{a}^{\rm{H}}\left(M_t, \theta_{p}\right),
\end{equation}
where $P$ is the total number of targets, while $b_{ p}$, $\theta_p$, and $\phi_p$ respectively denote the complex gain, the angle of arrival (AoA), and the angle of departure (AoD) of the $p$-th target. The array response vector is defined as $\mathbf{a}(M, \alpha)=[1,e^{-j2\pi d \frac{\sin (\alpha)}{\lambda }},{\cdots },e^{-j2\pi d (M-1) \frac{\sin (\alpha)}{\lambda }}]^{\rm{T}}$, with $d$ and $\lambda$ denoting the antenna spacing and the wavelength, respectively. In the Type II architecture, the superposition introduces severe mutual interference between the S\&C signals, making conventional CE methods ineffective. Intuitively, concurrent target sensing and CE introduce more channel parameters to be estimated compared to orthogonal transmission systems within Phase I. This problem is especially pronounced when the length of the UPs is insufficient. In such cases, this joint estimation problem with an insufficient number of UPs can also be considered as a variation of the pilot contamination problem \cite{5898372}. 

In summary, the distinction between Type I and Type II ISAC architectures fundamentally lies in the tradeoff between spectral efficiency and signal processing complexity. Specifically, the Type I system features lower signal processing complexity at the cost of extra time-frequency resources and dedicated scheduling between the CUs and the BS. In contrast, while the Type II system has superior spectral efficiency, it requires sophisticated receiver designs to counter the severe mutual interference between the UPs and the sensing signal. In this paper, we primarily focus on joint target sensing and communication signal detection in Phase II. Hence, the detailed joint estimation algorithm for Phase I is omitted, and it is assumed that perfect CSI is available at the receiver, i.e., $\hat{\mathbf{H}}_{\rm{c}}={\mathbf{H}}_{\rm{c}}$.



			\subsection{Phase II: Joint Target Sensing and Uplink Data Detection}
				
				The main motivation of uplink ISAC systems is to utilize the same frequency for the simultaneous transmission of the uplink communication signal and sensing signal to enhance spectral efficiency. Different from the pilots that occupy a small part of the frame, data transmission in Phase II typically occupies a long time period, which can largely improve the sensing performance if utilized. In the following, we discuss the S\&C signal models and formulate the joint target sensing and uplink data detection problem.

			The target sensing is implemented over a coherent processing interval consisting of $L$ snapshots. The transmitted sensing signal at the transmitter at the $l$-th snapshot,  ${\mathbf{x}}_{  {\rm{r}}}[l]\in \mathbb{C} ^{M_t\times 1}$, satisfies  $ \mathbb{E}\left[{\mathbf{x}}_{  {\rm{r}}}[l]\right]={\mathbf{0}} $ and $\mathbb{E}[\operatorname{Tr}( {{\mathbf{x}}_{  {\rm{r}}}[l]{{\mathbf{x}}^{\rm{H}}_{  {\rm{r}}} }}[l])] =\operatorname{Tr}(\mathbf{R})\le P_{ \rm{r}}$, where $\mathbf{R} \succeq 0$ and $P_{ \rm{r}}$ denote the covariance matrix and the maximum transmit power of the sensing signal, respectively. { Here, we assume that the receiver has no prior knowledge of the number of targets or their corresponding angles, and therefore the target response matrix estimation problem is considered \cite{9652071}{\footnote{{Target response matrix estimation is a simple linear estimation problem to ensure the optimality of the proposed receiver, which is widely considered in the existing literature \cite{10596930,8579200, OverviewZhang}. Moreover, the proposed framework can be extended to other estimation problems with minor modifications. }}}. After the target response matrix ${\mathbf{H}}_{ {\rm{r}}}$ is estimated, the angle information of the targets $\mathbf{\Theta}=\{\theta _p,\phi_p\}^{P}_{p=1}$ is extracted using existing signal processing methods, such as fast Fourier transform (FFT), multiple signal classification (MUSIC) \cite{10787076}, compressed sensing (CS)-based methods \cite{8827589}, and machine learning \cite{10496165}.} Denoting ${\mathbf{X}}_{ {\rm{r}}} = \left[ {{\mathbf{x}}_{ {\rm{r}}}[1], \cdots ,{\mathbf{x}}_{ {\rm{r}}}\left[ L \right]} \right] \in \mathbb{C} ^{M_t\times L}$, the average power of the received sensing signal across $L$ snapshots is given by $P_{\rm{s}}\triangleq \mathbb{E}[\|{\mathbf{H}}_{\rm{r}}{\mathbf{X}}_{  {\rm{r}}}\|_F^2]/L$. 
			

			At the $l$-th snapshot, the $k$-th CU transmits a modulated symbol $\tilde{{x}}_{ {\rm{c}},k} \in \mathcal{X}$ to the receiver for uplink communication. We assume that the transmitted symbols of $K$ CUs, ${{\tilde{\mathbf{x}}_{ {\rm{c}}}}}[l]=[\tilde{{x}}_{ {\rm{c}},1}[l],\cdots,\tilde{{x}}_{ {\rm{c}},K}[l]]^{\rm{T}}$, are mutually uncorrelated, satisfying $ \mathbb{E}\left[{{\tilde{\mathbf{x}}_{ {\rm{c}},k}}[l]}\right]={\mathbf{0}} $ and $\mathbb{E}\left[ {{{{\tilde{\mathbf{x}}_{ {\rm{c}}}}}[l]}{{{{\tilde{\mathbf{x}}_{ {\rm{c}}}}[l]}^{\rm{H}}} }}\right] = P_{\rm{c}} {\mathbf{I}}_K$, where $P_{\rm{c}}$ denotes the transmit power of the communication signal.  Due to the rich scattering environment, the channel matrix $\mathbf{H}_ {\rm{c}}$ has full rank, i.e., $\operatorname{Rank}(\mathbf{H}_{ {\rm{c}}})=K $, and the channel matrix is normalized so that $\mathbb{E}[\|{\mathbf{H}}_{ {\rm{c}}}\|_F^2]=K M_r$.

			We assume that the frequencies of the uplink communication signal and sensing signal are the same. Then, the received signal at the $l$-th  snapshot, ${\tilde{\mathbf{y}}}[l]\in \mathbb{C} ^{M_r\times 1}$, is a combination of the uplink communication signal and sensing echo, given by 
			\begin{align}
				{\tilde{\mathbf{y}}}[l]=&\underbrace{\mathbf{H}_{ {\rm{c}}} \tilde{\mathbf{x}}_{ {\rm{c}}}[l]}_{\textrm{Uplink signal}} +\underbrace{\mathbf{H}_{ \rm{r}} \mathbf{x}_{ \rm{r}}[l]}_{\textrm{Sensing echo}}+\tilde{\mathbf{n}}[l],
			\end{align}
			where $\tilde{\mathbf{n}}[l]$ denotes the additive white Gaussian noise (AWGN) at the receiver, with $\tilde{\mathbf{n}}[l] \sim {\mathcal C}{\mathcal N}\left( {{\mathbf{0}},\sigma^2{{\mathbf{I}}_{M_r}}} \right)$ and noise power $\sigma^2$.

			By stacking $L> M_t$ snapshots together, the received signal at the receiver can be formulated as 
			\begin{equation} \label{vdfz}
				{{\mathbf{Y}}} = {\mathbf{H}}_{ {\rm{r}}}{\mathbf{X}}_{ {\rm{r}}} + {\mathbf{H}}_{ {\rm{c}}}{\mathbf{X}}_{ {\rm{c}}} + {\tilde{\mathbf{N}}},
			\end{equation}
			where   ${\mathbf{X}}_{ {\rm{c}}} = \left[ {\tilde{\mathbf{x}}_{ {\rm{c}}}[1], \cdots ,\tilde{\mathbf{x}}_{ {\rm{c}}}\left[ L \right]} \right] \in \mathbb{C} ^{K\times L}$, ${\mathbf{Y}} =\left[ {	{\tilde{\mathbf{y}}}[1], \cdots ,	{\tilde{\mathbf{y}}}\left[ L \right]} \right] \in \mathbb{C} ^{M_r\times L}$, and  ${{\tilde{\mathbf{N}}}} =\left[ {{\tilde{\mathbf{n}}}[1], \cdots ,{{\tilde{\mathbf{n}}}}\left[ L \right]} \right] \in \mathbb{C} ^{M_r\times L}$. 
			
			In the considered uplink ISAC systems, the receiver is responsible for simultaneously extracting the target response matrix and decoding the communication signal. Since the AWGN at the receiver follows the distribution of $\mathbf{n}=\operatorname{vec}(\tilde{\mathbf{N}})\sim{\mathcal C}{\mathcal N}\left( {{\mathbf{0}}, \sigma^2{{\mathbf{I}}_{L M_r}}} \right)$, the probability density function of $\mathbf{Y}$ given ${\bf{X}}_{\rm{c}}$ and ${\bf{H}}_{\rm{r}}$ is
			\begin{equation} \label{vzz}
				p({\bf{Y}} \mid {\bf{X}}_{\rm{c}}, {\bf{H}}_{\rm{r}})=\frac{1}{(\pi \sigma^2)^{L M_r}} e^{\frac{-{\left\| {{\bf{Y}} -  {\mathbf{H}}_{ {\rm{r}}}{\mathbf{X}}_{ {\rm{r}}} - {\mathbf{H}}_{ {\rm{c}}}{\mathbf{X}}_{ {\rm{c}}}} \right\|_F^2}}{\sigma^2}  }.
			\end{equation}
			Thus, the ML estimator can be formulated as a mixed-integer LS problem
			\begin{equation} \label{eqvvq}
				\mathop {\operatorname{argmin} }\limits_{{\bf{H}}_{\rm{r}},\,{{\bf{X}}_{\rm{c}}}\in  \mathcal{X}^{ K \times L}}{\left\| {{\bf{Y}} -  {\mathbf{H}}_{ {\rm{r}}}{\mathbf{X}}_{ {\rm{r}}} - {\mathbf{H}}_{ {\rm{c}}}{\mathbf{X}}_{ {\rm{c}}}} \right\|_F^2}   .
			\end{equation}
			According to the equality $\operatorname{vec}(\mathbf{A} \mathbf{C})=\left(\mathbf{I} \otimes \mathbf{A}\right) \operatorname{vec}(\mathbf{C})=\left(\mathbf{C}^{\mathrm{T}} \otimes \mathbf{I}\right) \operatorname{vec}(\mathbf{A})$ \cite{zhang2017matrix},  (\ref{eqvvq}) can be reformulated as
			\begin{equation} \label{eqq}
				\mathop {\operatorname{argmin} }\limits_{{\bf{h}}_{\rm{r}},\,{{\bf{x}}_{\rm{c}}}\in  \mathcal{X}^{LK}}{\left\| {{\bf{y}} - {\bf{A}}_{\rm{c}}{\bf{x}}_{\rm{c}} - {\bf{A}}_{\rm{r}}{\bf{h}}_{\rm{r}}} \right\|_2^2}   ,
			\end{equation}
			where ${\bf{x}}_{\rm{c}}=\operatorname{vec}(\mathbf{X}_{\rm{c}}) \in \mathbb{C} ^{LK\times 1}$,   ${\mathbf{y}}=\operatorname{vec}(\mathbf{Y}) \in \mathbb{C} ^{LM_r \times 1} $, ${\bf{A}}_{\rm{r}} = {\bf{X}}_{\rm{r}}^{\rm{T}} \otimes {\bf{I}}_{M_r} \in \mathbb{C} ^{L M_r\times M_r M_t}$,  ${\bf{A}}_{\rm{c}} = {\bf{I}}_{L} \otimes {\bf{H}}_{\rm{c}} \in \mathbb{C} ^{L M_r\times L K}$, and the target response vector is given by ${\bf{h}}_{\rm{r}}=\operatorname{vec}(\mathbf{H}_{\rm{r}}) \in \mathbb{C} ^{M_r M_t\times 1}$. 
			
			\begin{remark}
				Retrieving the communication signal and the target response matrix in Problem (\ref{eqvvq}) is both fundamental and challenging. Different from conventional linear LS problems that have a unique solution if the observation matrix has full column rank, solving the equation $\mathbf{AX}+\mathbf{ZB}=\mathbf{C}$ naturally provides a non-unique solution set unless other constraints are imposed \cite{Liao2005best}. In Problem (\ref{eqvvq}), the uniqueness is guaranteed by considering the communication alphabet constraint, which is further investigated in the subsequent sections.
			\end{remark}

			\section{Related Work}
			\begin{figure}
				\centering
				\includegraphics[width=8.5cm]{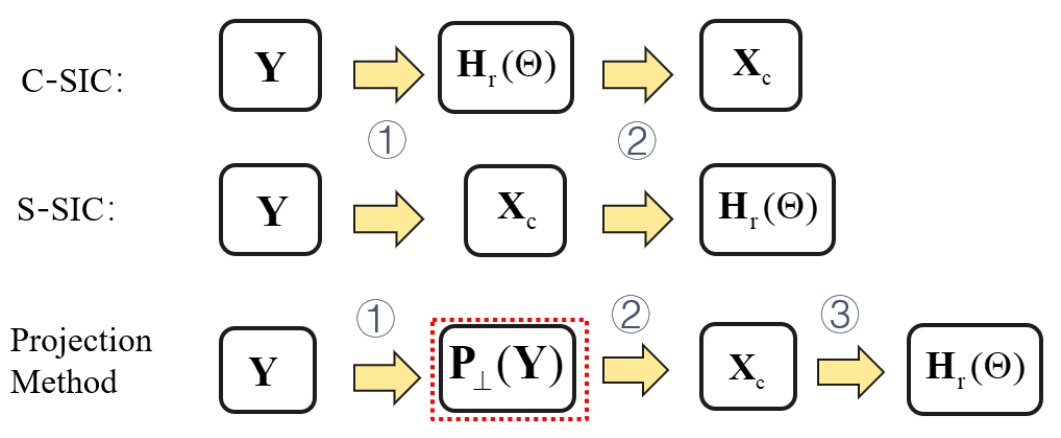} 
				\caption{\small{Various types of receiver designs in uplink ISAC systems}}
				\vspace{-0.5cm}
			\end{figure}
			
			\subsection{The SIC-Type Receivers}
			Previous contributions mainly considered two types of SIC-based schemes to address Problem (\ref{eqvvq}), namely communication-oriented SIC (C-SIC) and sensing-oriented SIC (S-SIC) \cite{SICOuyang}. As shown at the top of Fig. 3, C-SIC and S-SIC both use two stages to process the S\&C signal, where the estimated signal in the first stage is subtracted from the superposed signal in the second stage. In the S-SIC scheme, the receiver first decodes the uplink data by treating the aggregate interference-plus-noise ${\bf{A}}_{\rm{r}}{\bf{h}}_{\rm{r}}+\mathbf{n}$ as  Gaussian noise,  and the decoded signal can be obtained by solving the following standard signal detection problem
			\begin{equation} \label{zbvv}
				{\bf{x}}_{\rm{c}}^{\rm{SIC}}=\mathop {\operatorname{argmin} }\limits_{{{\bf{x}}_{\rm{c}} \in  \mathcal{X}^{LK}}}{\left\| {{\bf{y}} - {\bf{A}}_{\rm{c}}{\bf{x}}_{\rm{c}} } \right\|^2_2 }  .
			\end{equation}
			Assuming that the uplink data has been perfectly decoded, the detected communication signal is subtracted from the superposed signal. The remainder, ${{\bf{y}} - {\bf{A}}_{\rm{c}}{\bf{x}}_{\rm{c}}^{\rm{SIC}} }$, is then used to estimate the target response matrix ${\bf{H}}_{\rm{r}}$ by the standard LS method. When all the communication symbols are successfully detected, the target response estimation can achieve the same performance as the sensing-only system. 
			
			{In the C-SIC scheme, the target response vector is first estimated under the communication interference, and the estimated radar signal is subtracted during communication signal detection. However, the residual target response estimation error influences the subsequent communication performance, indicating that the S\&C performance of the C-SIC scheme in two stages is strictly worse than that of the sensing-only and communication-only systems. To fully exploit the potential of uplink ISAC systems,  the S-SIC scheme (which is called the SIC scheme in the rest of the paper) is mainly discussed throughout the rest of this paper.}
			
			%
			
			The most significant drawback of the SIC scheme in uplink ISAC systems is the ineffective interference processing function. By treating the radar signal as Gaussian noise, the communication signal detection problem in (\ref{zbvv}) has a low SINR, leading to a high BER in the first stage. Furthermore, the residual communication signal detection error becomes more pronounced due to the high BER in the first stage, significantly impacting the target response estimation performance in the second stage, known as error propagation.
			
			\vspace{-0.2cm}
			\subsection{The Projection-Type Receiver}

			\begin{theorem}
				The solutions to Problem (\ref{eqvvq}) are equivalent to the combination of the solution for the transformed signal detection Problem (\ref{vbdf}) 
			\begin{equation} \label{vbdf}
				{\hat{ \mathbf{x}}_{\rm{c}}}=\mathop {\operatorname{argmin}}\limits_{{{\mathbf{x}}_{\mathrm{c}}} \in  \mathcal{X}^{LK}}  \quad \| \mathbf{\Gamma}(\mathbf{y}-\mathbf{A}_{\rm{c}}{\mathbf{x}}_{\mathrm{c}})\|_2^2,
			\end{equation}
			and the closed-form solution for the target response vector with the estimated communication symbols ${\hat{ \mathbf{x}}_{\rm{c}}}$: 
			\begin{equation} \label{bav}
				\hat{\mathbf{h}}_{\rm{r}}({\hat{ \mathbf{x}}_{\rm{c}}})=\mathbf{\Xi} \mathbf{A}^{\rm{H}}_{{\rm{r}}}(\mathbf{y}-\mathbf{A}_{\rm{c}}{\hat{ \mathbf{x}}_{\rm{c}}}),
			\end{equation}
			where $\mathbf{\Xi}=\left(\mathbf{A}_{\rm{r}}^{\mathrm{H}} \mathbf{A}_{\rm{r}		}\right)^{-1}$, and $\mathbf{\Gamma}=\mathbf{I}_{{LM}_r}-\mathbf{A}_{\rm{r}}(\mathbf{A}_{\rm{r}}^{\rm{H}}\mathbf{A}_{\rm{r}})^{-1}\mathbf{A}_{\rm{r}}^{\rm{H}}$ is the projection matrix of the complement space of $\mathbf{A}_{\rm{r}}$.
			\end{theorem}
			
			\begin{proof}
				Please refer to \cite{10663294}.
			\end{proof}
			
			Based on Theorem 1, the projection-based algorithm is provided in \cite[Algorithm 1]{10663294} and at the bottom of Fig. 2. Our proposed projection-type receiver can be regarded as an extension to the SIC scheme with proper interference elimination methodology.  Compared with the SIC scheme, we first project the superimposed signal into the complement space of $\mathbf{A}_{\rm{r}}$ to eliminate the radar signal interference in the signal detection process. After this projection, we decode the communication signal and subsequently estimate the target response, following a process similar to the SIC method.

				
					%
			{It was demonstrated in \cite{10663294} that the proposed projection-type receiver maintains the same signal-to-noise ratio (SNR) as the communication-only systems in the signal detection problem and has lower BER than conventional SIC in most cases.} Moreover, jointly processing multiple snapshots is another key distinction between the projection-type receiver and the SIC-type receiver. It was shown that the signal detection performance improves and the communication ergodic achievable rate increases as the number of snapshots $L$ increases. As $L\rightarrow \infty$, the ergodic rate approaches that of the communication-only systems, indicating that the impact of the mutual interference can be perfectly canceled.
			
			However, the improved performance comes at the cost of increased complexity. The projection-type receiver requires handling a high-dimensional equivalent channel matrix $\mathbf{G}\triangleq \mathbf{\Gamma}\mathbf{A}_{\rm{c}} \in \mathbb{C} ^{L M_r\times LK}$ in (\ref{vbdf}) that scales with the product of the number of transmitted antennas and the number of snapshots.
			In addition to the substantially increased dimension, the performance of the projection-type receiver also suffers from the rank-deficiency issue, i.e., the rank of the equivalent channel matrix $\mathbf{G}$ is smaller than the number of transmitted symbols from all users. This rank-deficiency issue is presented in the following lemma.
			
			\begin{lemma}
				Matrix $\mathbf{G}$ is a singular matrix, and its rank is given by $\operatorname{Rank} (\mathbf{G})=(L-M_t)K< L K$.
			\end{lemma}

			\begin{proof}
				Please refer to \cite{10663294}.
			\end{proof}
			
			\begin{remark}
				{ From a matrix equation perspective, solving a rank-deficient matrix equation leads to infinite solutions, which are extremely sensitive to noise.} Since the projection-type receiver establishes an equivalent transformation of Problem (\ref{eqvvq}), it also maintains the rank deficiency issue, posing a challenge in the communication signal detection problem.
			\end{remark}

				\vspace{-0cm}
				\section{The FP-type Receiver: A Unified Framework}
				{In the projection-type receiver, the received signal is projected to the complement space of the radar signal so that the radar interference is fully eliminated during signal detection. In this section, we first introduce a more general receiver design principle, called the FP-type receiver. The FP-type receiver unifies the aforementioned projection-type receiver and the SIC-type receiver with a flexible tradeoff factor.} Then, the SINR and the condition number of the formulated signal detection problem using the FP-type receiver are explored. Finally, the PEPs of the FP-type receiver using the ML detector and a linear detector are analyzed, respectively.
				\subsection{Design Principle of the FP-type Receiver}
				The core idea of the projection-type receiver is to solve the signal detection problem in (\ref{vbdf}). In the projection-type receiver, matrix $\mathbf{\Gamma}$ is the orthogonal projection matrix of $\mathbf{A}_{\rm{r}}$, which can be rewritten as
				\begin{equation} \label{brr}
					\mathbf{\Gamma}=\mathbf{P}_{\perp} \otimes \mathbf{I}_{M_r},
				\end{equation}
				and the equivalent channel matrix in the signal detection problem (\ref{vbdf}) is 
				\begin{equation} \label{br}
					\mathbf{G}=\mathbf{\Gamma}\mathbf{A}_{\rm{c}}=\mathbf{P}_{\perp} \otimes \mathbf{H}_{\rm{c}},
				\end{equation}
				where $\mathbf{P}_{\perp }=\mathbf{I}_L-{\bf{X}}_{\rm{r}}^{\rm{T}}({\bf{X}}_{\rm{r}}^{*}{\bf{X}}_{\rm{r}}^{\rm{T}})^{-1}{\bf{X}}_{\rm{r}}^{*}$ is the projection matrix corresponding to the complement space of the radar signal.  
				
				It can be observed from (\ref{brr}) and (\ref{br}) that the formulated signal detection problem in (\ref{vbdf}) can be modified by changing matrix $\mathbf{P}_{\perp }$. Moreover, the singular values of $	\mathbf{G}$ are also determined by those of the orthogonal projection matrix $\mathbf{P}_{\perp}$, which introduces the rank deficiency issue. Hence, modifying matrix $\mathbf{P}_{\perp }$ can significantly influence the properties of the transformed signal detection problem.

				{Motivated by the above observation, we define $\mathbf{P}_{\parallel }={\bf{X}}_{\rm{r}}^{\rm{T}}({\bf{X}}_{\rm{r}}^{*}{\bf{X}}_{\rm{r}}^{\rm{T}})^{-1}{\bf{X}}_{\rm{r}}^{*}$ as the  projection matrix corresponding to the signal space of the radar waveform. Then, the projection matrix of the FP-type receiver  $\mathbf{P}_{ \rm{FP}}$ is updated according to
					\begin{equation} \label{bbdf}
						\begin{aligned}
							\mathbf{P}_{ \rm{FP}}&\triangleq(1-\rho)\mathbf{P}_{\perp}+\rho \mathbf{I}=\mathbf{P}_{\perp}+\rho \mathbf{P}_{\parallel }, 
						\end{aligned}
					\end{equation}
					where $\rho \in [0,1]$ is the tradeoff factor.} Upon defining $\mathbf{\Gamma}_{\parallel}=\mathbf{P}_{\parallel} \otimes \mathbf{I}_{M_r}$ as the projection matrix corresponding to the signal space of $\mathbf{A}_{\rm{r}}$ and $\mathbf{\Gamma}_{\rm{FP}}=\mathbf{P}_{\rm{FP}} \otimes \mathbf{I}_{M_r}$, the communication signal detection in the FP-type receiver can be formulated as 
				\begin{small}
					\begin{align} \label{xc}
						{\hat{ \mathbf{x}}_{\rm{c}}}=&\mathop {\operatorname{argmin}}\limits_{{{\mathbf{x}}_{\mathrm{c}}} \in  \mathcal{X}^{LK}}  \quad \| \mathbf{\Gamma}_{\rm{FP}}(\mathbf{y}-\mathbf{A}_{\rm{c}}{\mathbf{x}}_{\mathrm{c}}-\mathbf{A}_{\rm{r}}{\mathbf{h}}_{\mathrm{r}})\|_2^2 \\
						=&\mathop {\operatorname{argmin}}\limits_{{{\mathbf{x}}_{\mathrm{c}}} \in  \mathcal{X}^{LK}}   \| \underbrace{\mathbf{\Gamma}(\mathbf{y}-\mathbf{A}_{\rm{c}}{\mathbf{x}}_{\mathrm{c}})}_{
							\begin{aligned}
								\text{Complement Space} \\
								\text{Component}
						\end{aligned}} 
						+\rho \underbrace{\mathbf{\Gamma}_{\parallel}(\mathbf{y}-\mathbf{A}_{\rm{c}}{\mathbf{x}}_{\mathrm{c}}-\mathbf{A}_{\rm{r}}{\mathbf{h}}_{\mathrm{r}})}_{
							\begin{aligned}
								\text{Signal Space} \\
								\text{Component}
						\end{aligned}} \|_2^2. \nonumber
					\end{align}
				\end{small}
				\begin{figure}
					\centering
					\begin{minipage}[t]{0.49\textwidth}
						\centering
						\includegraphics[width=9cm]{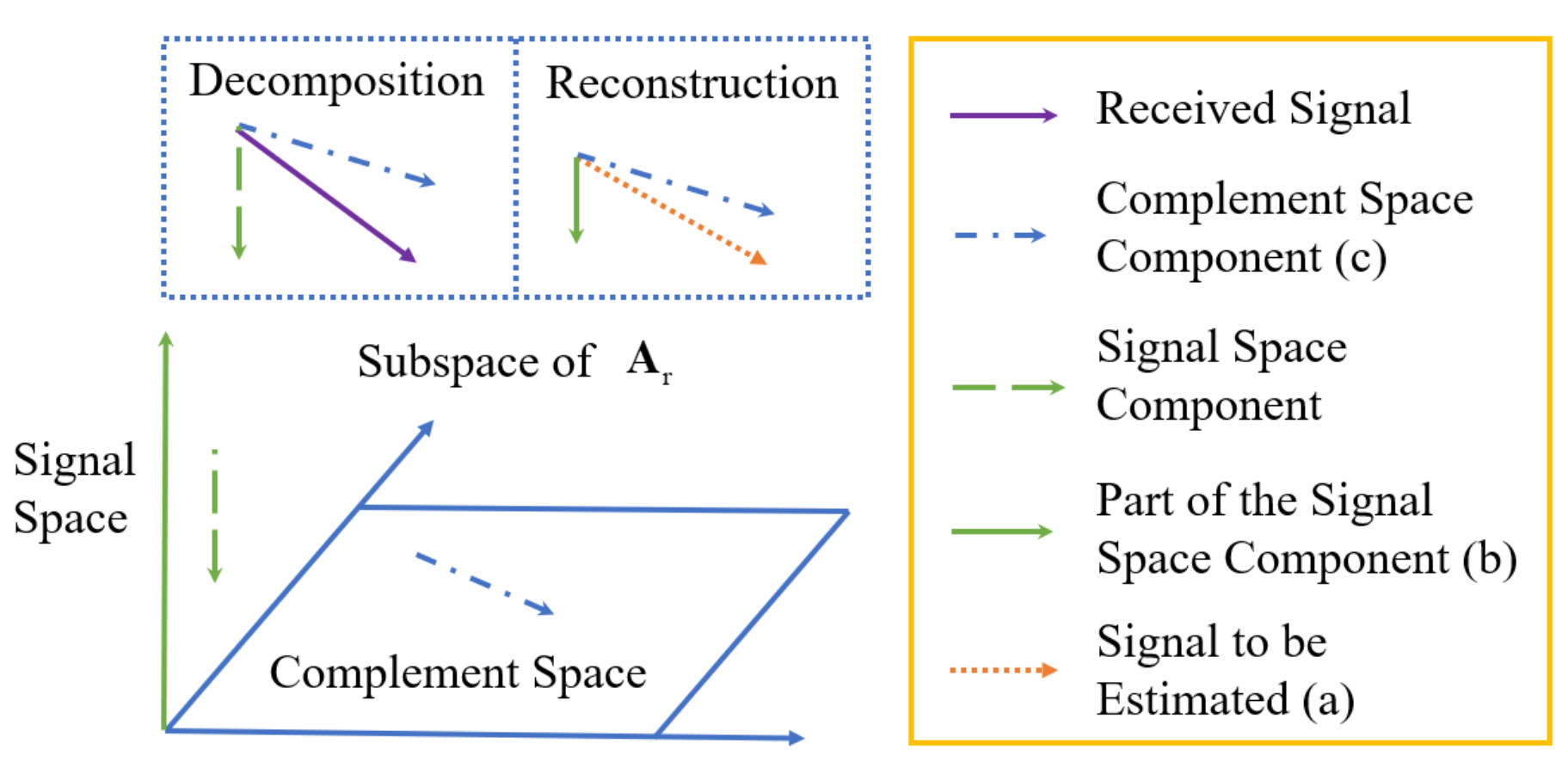}
						\caption{\small{Illustration of the proposed FP-type receiver}}
					\end{minipage}
				\end{figure}
				
				In Fig. 4, we provide an illustration of the proposed FP-type receiver.  The received signal is decomposed into the signal space component and the complement space component of matrix $\mathbf{A}_{\rm{r}}$, respectively. In the proposed FP-type receiver, the communication signal is detected via a reconstructed signal, as depicted in Fig. 4(a), which is the combination of the complement space component, as depicted in Fig. 4(b), and part of the signal space component, as depicted in Fig. 4(c). The tradeoff factor $\rho$ decides the ratio of the signal space component used.  More specifically, if the tradeoff factor $\rho=1$, we have $\mathbf{P}_{ \rm{FP}}=\mathbf{I}_{M_r}$ and $\mathbf{\Gamma}_{ \rm{FP}}=\mathbf{I}_{LM_r}$. This means the reconstructed signal is the same as the received signal, which corresponds to the SIC-type receiver. By decreasing the tradeoff factor $\rho$,  the signal space component has less impact on determining the communication symbols. When $\rho=0$, the received signal is projected to the complement space of the radar signal to fully eliminate the radar interference before the signal detection. In other words, the FP-type receiver reduces to the projection-type receiver. 
				
				Since we have no prior information about $\mathbf{h}_{\rm{r}}$, a similar methodology as in the SIC-type receiver is used by treating the signal space component of the radar signal plus noise as the AWGN. Then, Problem (\ref{xc}) can be reformulated as 
				\begin{equation} \label{fbgfbg}
					\begin{aligned}
						{\hat{ \mathbf{x}}_{\rm{c}}}\approx &\mathop {\operatorname{argmin}}\limits_{{{\mathbf{x}}_{\mathrm{c}}} \in  \mathcal{X}^{LK}}  \quad \| \mathbf{\Gamma}_{\rm{FP}}(\mathbf{y}-\mathbf{A}_{\rm{c}}{\mathbf{x}}_{\mathrm{c}})\|_2^2  \\
						=&\mathop {\operatorname{argmin}}\limits_{{{\mathbf{x}}_{\mathrm{c}}} \in  \mathcal{X}^{LK}}  \quad \| \mathbf{y}_{\rm{FP}}-\mathbf{G}_{\rm{FP}}{\mathbf{x}}_{\mathrm{c}}\|_2^2,
					\end{aligned}
				\end{equation}
				where 
				\begin{align} \label{ngf}
					\mathbf{y}_{\rm{FP}}=(\mathbf{P}_{ \rm{FP}} \otimes \mathbf{I}_{M_r}) \mathbf{y}, \;\text{and} \; 
					\mathbf{G}_{\rm{FP}}=\mathbf{P}_{ \rm{FP}} \otimes \mathbf{H}_{\rm{c}}.
				\end{align}
				
				After the solution to Problem (\ref{fbgfbg}) is obtained, the FP-type receiver then performs the target response estimation with the estimated communication signal ${\hat{ \mathbf{x}}_{\rm{c}}}$ according to (\ref{bav}). Since the precision of the target response estimation is largely determined by the performance of the signal detection, the performance of the FP-type receiver is evaluated by the characteristics of Problem (\ref{fbgfbg}) in the subsequent analysis.

				\subsection{Revealing the Tradeoff Between Condition Number and SINR}
				In this subsection, we analyze the condition number and the SINR of the transformed signal detection problem using the FP-type receiver.
				
				{The condition number of matrix $\mathbf{G}_{\rm{FP}}$ characterizes the correlation between each column of the matrix, given by 
					\begin{equation}
						\begin{aligned}
							\textrm{Cond}(\mathbf{G}_{\rm{FP}})&\triangleq  { \frac{\omega_\textrm{1}(\mathbf{G}_{\rm{FP}}) }{\omega_{LK}(\mathbf{G}_{\rm{FP}})} } = \sqrt{ \frac{\lambda_\textrm{1}(\mathbf{G}_{\rm{FP}}^{\rm{H}}\mathbf{G}_{\rm{FP}}) }{\lambda_{LK}(\mathbf{G}_{\rm{FP}}^{\rm{H}}\mathbf{G}_{\rm{FP}})} } \\
							&=\sqrt{\frac{\lambda_\textrm{1}(\mathbf{P}_{ \rm{FP}}^{\rm{H}}\mathbf{P}_{ \rm{FP}})\lambda_\textrm{1}(\mathbf{H}_{\rm{c}}^{\rm{H}}\mathbf{H}_{\rm{c}}) }{\lambda_{L}(\mathbf{P}_{ \rm{FP}}^{\rm{H}} \mathbf{P}_{ \rm{FP}}) \lambda_{K}(\mathbf{H}_{\rm{c}}^{\rm{H}}\mathbf{H}_{\rm{c}})} } \\
							&=\frac{\lambda_\textrm{1}(\mathbf{P}_{ \rm{FP}}) }{\lambda_{L}(\mathbf{P}_{ \rm{FP}})}\textrm{Cond}(\mathbf{H}_{\rm{c}})\overset{(a)}{=}\frac{1 }{\rho}\textrm{Cond}(\mathbf{H}_{\rm{c}}),
						\end{aligned}
					\end{equation}
					where (a) exploits the fact that $\mathbf{P}_{\perp}$ has $L-M_t$ unit eigenvalues as well as $M_t$ zero eigenvalues, and for arbitrary matrix $\mathbf{A} \in \mathbb{C} ^{N \times N} $, $\lambda_{i}(\mathbf{A}+a \mathbf{I})=\lambda_{i}(\mathbf{A})+a, i=1,\cdots, N$.}
				
				The above result indicates that with a smaller tradeoff factor, the column correlation increases, leading to an ill-conditioned equivalent channel matrix and worse signal detection performance.

				\begin{lemma}
					The SINR of the transformed signal detection problem using the FP-type receiver is given by 
				\begin{equation} \label{vdfbvb}
					\begin{aligned}
						\textrm{SINR}_{\rm{FP}}&\triangleq \frac{\mathbb{E}\left[ {\|\mathbf{\Gamma}_{\rm{FP}} \mathbf{A}_{\rm{c}}\mathbf{x}_{\rm{c}}\|_2^2}\right] }{\mathbb{E}\left[ {\|\mathbf{\Gamma}_{\rm{FP}} \mathbf{A}_{\rm{r}}\mathbf{h}_{\rm{r}}\|_2^2}\right] + \mathbb{E}\left[  \|\mathbf{\Gamma}_{\rm{FP}}\mathbf{n}\|_2^2\right]} \\
						&= \frac{P_{\rm{c}}(L-(1-\rho^2)M_t) K M_r}{\rho^2 L P_{\rm{s}}+(L-(1-\rho^2)M_t)M_r  \sigma^2},
					\end{aligned}
				\end{equation}
				which is a monotonically decreasing function of the tradeoff factor $\rho$. The maximum and minimum SINRs are achieved by the projection-type and SIC-type receivers, respectively.
				\end{lemma}
				
				
				\begin{proof}
					Please refer to Appendix A.
				\end{proof}
				
				\begin{remark}
					From the geometry perspective, the received signal that lies in the complement space has a higher SINR as compared to the signal space component. However, directly projecting the received signal to the complement space reduces the dimension of the observation signal, leading to a correlated equivalent channel matrix. By increasing the tradeoff factor $\rho$, the channel correlation is decreased owing to the introduced signal space component,  which, however, also leads to a decreased SINR.
				\end{remark}

				\subsection{PEP Analysis of the FP-type Receiver}
				The condition number and SINR of the signal detection problem derived in the last subsection characterize the properties of the formulated problem from different aspects. The  error probability $P_e$ under different algorithms is a more direct performance metric, which however is generally computationally intractable. A customary solution is to approximate the error probability using the PEP.  The PEP is defined as the probability that $\bar{\mathbf{x}}_{\rm{c}}$ is selected at the receiver when $\tilde{\mathbf{x}}_{\rm{c}}$ is transmitted, i.e., $	P(\tilde{\mathbf{x}}_{\rm{c}} \to \bar{\mathbf{x}}_{\rm{c}})$. The connection between the error probability and PEP can be characterized by
				\begin{equation}
					P_e \leq \frac{1}{|\mathcal{X}|} \sum_{\tilde{\mathbf{x}}_{\rm{c}} \in \mathcal{X}} \sum_{\bar{\mathbf{x}}_{\rm{c}} \in \mathcal{X} \setminus \tilde{\mathbf{x}}_{\rm{c}}} P(\tilde{\mathbf{x}}_{\rm{c}} \to \bar{\mathbf{x}}_{\rm{c}}).
				\end{equation}
				In what follows, we analyze the PEP of the FP-type receiver when using the ML detector and the ZF detector, respectively.
				\subsubsection{ML Detector}
				ML detector uses an exhaustive searching method to find the solution to Problem (\ref{fbgfbg}) and typically serves as the performance upper bound of the signal detection algorithm. Denoting $\boldsymbol{\delta }=\tilde{\mathbf{x}}_{\rm{c}} - \bar{\mathbf{x}}_{\rm{c}}$,
				the PEP under the ML criterion is given by\cite[Chapter 15]{proakis2008digital}
				\begin{align} \label{bx}
					P_{\rm{ML}}(\tilde{\mathbf{x}}_{\rm{c}} \to \bar{\mathbf{x}}_{\rm{c}}) &= P\left( \|\tilde{\mathbf{y}} - \mathbf{G}_{\rm{FP}}\bar{\mathbf{x}}_{\rm{c}}\|_2 \leq \|\tilde{\mathbf{y}} - \mathbf{G}_{\rm{FP}}\tilde{\mathbf{x}}_{\rm{c}}\|_2 \right) \nonumber \\
					&\overset{(a)}{\approx}  Q\left( \frac{\|\mathbf{G}_{\rm{FP}}\boldsymbol{\delta }\|_2}{\sqrt{2\sigma_{\rm{ML}}^2}} \right),
				\end{align}
			where $Q$ denotes the Q-function and $(a)$ assumes the residual interference plus noise $\hat{\mathbf{n}}\triangleq \tilde{\mathbf{y}} - \mathbf{G}_{\rm{FP}}\tilde{\mathbf{x}}_{\rm{c}}$ is spatially white with the normalized equivalent noise power  
				\begin{equation}
					\begin{aligned}
						\sigma_{\rm{ML}}^2 &= \frac{\rho^2 \mathbb{E} \left[ {\| \mathbf{A}_{\rm{r}}\mathbf{h}_{\rm{r}}\|_2^2}\right]+ \mathbb{E}\left[ {\|\mathbf{\Gamma}_{\rm{FP}}\mathbf{n}\|_2^2}\right]}{K P_{\rm{c}}} \\
						&= \frac{\rho^2 L P_{\rm{s}} + (L - (1 - \rho^2)M_t) M_r \sigma^2}{K P_{\rm{c}}}.
					\end{aligned}
				\end{equation}
				Defining $\boldsymbol{\Delta }=\operatorname{Unvec}(\boldsymbol{\delta })\in  \mathbb{C} ^{K \times L}$ and using the equation $\operatorname{vec}(\mathbf{A}\mathbf{B}\mathbf{C}) = (\mathbf{C}^{\rm{T}} \otimes \mathbf{A})\operatorname{vec}(\mathbf{B})$, (\ref{bx}) can be rewritten as
				\begin{equation}
					P_{\rm{ML}}(\tilde{\mathbf{x}}_{\rm{c}} \to \bar{\mathbf{x}}_{\rm{c}})=Q\left( \frac{\|\mathbf{H}_{\rm{c}}\boldsymbol{\Delta }\mathbf{P}_{\rm{FP}}^{*}\|_F}{\sqrt{2\sigma_{\rm{ML}}^2}} \right).
				\end{equation}
				
				To derive the analytical results and provide guidance to the FP-type receiver design, the following two assumptions are made. 
				
				\itshape \textbf{Assumption 1:} \upshape Assume that the number of received antennas at the receiver goes to infinity, i.e., $M_r \to \infty $, and each entry of the channel matrix follows the complex Gaussian distribution, we have $\mathbf{H}_{\rm{c}}^{\rm{H}}\mathbf{H}_{\rm{c}} \approx \mathbf{I}_{K}$. This assumption is known as the favorable propagation in large-scale MIMO systems and is widely used in reducing the complexity of the MIMO precoding and signal detection\cite{6798744}. The favorable propagation assumption may not necessarily hold true in practical systems. However, adopting this assumption provides a theoretical upper bound of the PEP for ML detection and helps derive the analytical results. 
				
				\itshape \textbf{Assumption 2:} \upshape {The transmission process is defined as transmitting $L$ snapshots at a time block by $K$ CUs. We assume that during each transmission, at most only one transmitted symbol is incorrectly detected.} This assumption holds when the SINR is not very small and the number of snapshots is not very large{\footnote{{Assuming that symbol errors are independent and identically distributed (i.i.d.) with a symbol error rate (SER)  of $p$, the number of erroneous symbols in a data block, denoted as $X$, follows a binomial distribution $X \sim \text{Binomial}(N, p)$. For ML detection with typical parameters $K=8$, $L=16$ (i.e., block length $N=128$), and $p=10^{-3}$, the probability of exactly one error is $P(X=1) \approx 0.1127$, whereas the probability of more than one error is merely $P(X>1) \approx 0.0075$. This demonstrates that the single-symbol error event significantly dominates when an error occurs.  }}}.
			 Without loss of generality, we assume that only the transmitted symbol at the $K_{1}$-th CU in the $L_1$-th snapshot is incorrectly detected, then the $(p,q) $-th element of $\boldsymbol{\Delta }$ follows
				\begin{equation}
					\boldsymbol{\Delta }_{[p,q]}=\left\{\begin{array}{ll}
						d_{\rm{min}}, & p=K_{1},q=L_{1} \\
						0, & \textrm{else,}
					\end{array}\right.
				\end{equation}
				where $d_{\rm{min}}$ is the minimum distance in the constellation.
				
The instantaneous PEP highly depends on specific transmission conditions and error patterns, such as the exact error locations within the error matrix $\boldsymbol{\Delta }$ in ML detection, or the specific decoupled space-time subchannels in ZF detection. To derive a tractable closed-form analytical bound that can guide the receiver design, it is necessary to evaluate the expected or average PEP. By taking the expectation over the error matrix $\boldsymbol{\Delta }$ in ML detection, or averaging the error probabilities across all decoupled subchannels in ZF detection, we can capture how core parameters, such as the tradeoff factor $\rho$, impact the overall signal detection performance.
				
	\begin{lemma}
		By using the above two assumptions, the expectation of $P_{\rm{ML}}(\tilde{\mathbf{x}}_{\rm{c}} \to \bar{\mathbf{x}}_{\rm{c}})$ over $\boldsymbol{\Delta }$ can be approximated as  
		\begin{equation} \label{vsd}
			\begin{aligned}
				\mathbb{E}\left[P_{\rm{ML}}(\tilde{\mathbf{x}}_{\rm{c}} \to \bar{\mathbf{x}}_{\rm{c}})\right] {\approx} 
				 Q\left( \frac{ d_{\rm{min}}\sqrt{L-(1-\rho^2)M_t}}{ \sqrt{2L \sigma_{\rm{ML}}^2}} \right).
			\end{aligned}
		\end{equation}
	\end{lemma}

	\begin{proof}
		Please refer to Appendix C.
	\end{proof}
		
	It can be easily verified that for the FP-type receiver, the approximate PEP is an increasing function of $\rho$, indicating that the projection-type receiver is optimal if the ML detection is used.	

\subsubsection{ZF Detector}The performance of the linear estimator usually serves as the lower bound of the system performance. For linear reception, the ZF detector achieves complete cancellation of multi-user interference (MUI) by left-multiplying the post-processed received signal $\tilde{\mathbf{y}}$ with the Moore-Penrose pseudo-inverse of the equivalent channel matrix $\mathbf{G}_{\rm{FP}}$. The output of the ZF detector is formulated as
\begin{equation} \label{nfnnf}
\mathbf{\hat{x}}_{\rm{c}} = \mathbf{G}_{\rm{FP}}^{\dagger}\tilde{\mathbf{y}} = \mathbf{x}_{\rm{c}} + \mathbf{G}_{\rm{FP}}^{\dagger}\hat{\mathbf{n}},
\end{equation}
where $\mathbf{G}_{\rm{FP}}^{\dagger} = (\mathbf{G}_{\rm{FP}}^{\rm{H}}\mathbf{G}_{\rm{FP}})^{-1}\mathbf{G}_{\rm{FP}}^{\rm{H}}$. Then, its average PEP $\mathbb{E}\left[P_{\rm{ZF}}(\tilde{\mathbf{x}}_{\rm{c}} \to \bar{\mathbf{x}}_{\rm{c}})\right]$ is provided in the following Lemma.  

\begin{lemma}
By using Assumption 1, the average PEP of the ZF detector across all spatial-temporal symbols can be approximated as
\begin{align}  \label{bfg}
	\mathbb{E}\left[P_{\rm{ZF}}(\tilde{\mathbf{x}}_{\rm{c}} \to \bar{\mathbf{x}}_{\rm{c}})\right] =Q\left( \frac{	d_{\rm{min}} \sqrt{L} }{ \sqrt{2\sigma_{\rm{ML}}^2 (L-M_t+\frac{M_t}{\rho^2})}} \right).  
\end{align}
\end{lemma}

\begin{proof}
Please refer to Appendix B.
\end{proof}

This means that the detection performance of the ZF detector is determined not only by the equivalent noise of the ML detector $\sigma_{\rm{ML}}^2$, but also by the ``noise amplification'' effect of the pseudo-inverse of matrix $ \mathbf{P}_{\rm{FP}}$. When $\rho \to 0$, the matrix $ \mathbf{P}_{\rm{FP}}$ is reduced to that of the projection-type receiver, which theoretically has infinite noise. Therefore, conventional projection-type receiver is not suitable for the ZF detector.

				\begin{figure}
					\centering
					\includegraphics[width=8.5cm]{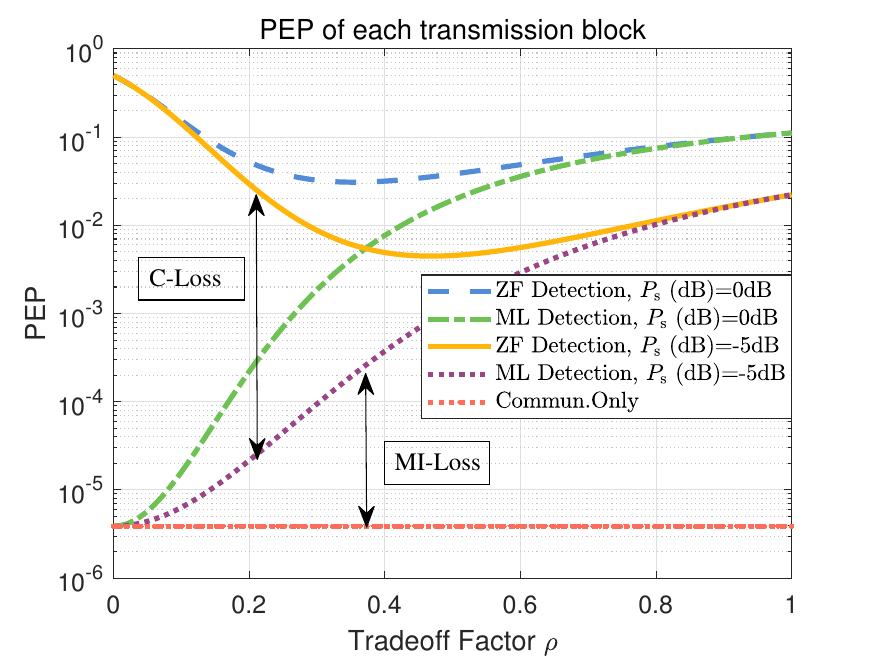} 
					\caption{\small{PEP of ZF and ML detectors with different sensing powers}}
				\end{figure}

				{ In Fig. 5, we examine the PEP of the ZF and ML detectors derived in (\ref{vsd}) and (\ref{bfg}). The transmit power of the communication signal is set to $P_{\rm{c}}=1$ W and the noise power is set to $\sigma^2=-10$ dBW. In the following, two observations on the performance loss and the selection of the receiver are discussed.


					\begin{remark}
						First, the PEPs of the ZF and ML detectors in communication-only systems are essentially the same with an infinite number of receive antennas due to favorable propagation \cite{1013126}. However, in the considered uplink ISAC systems, the performance gap between ZF and ML detection is non-negligible due to the processing of the sensing interference. This performance gap, termed the C-Loss, is a decreasing function with respect to the tradeoff factor and is close to zero when the tradeoff factor equals one. Second, the PEP of ML detection for uplink ISAC becomes larger than that of communication-only systems, and the corresponding performance gap is termed the MI-Loss. As observed in Fig. 5, the MI-Loss is {an increasing function} with respect to the tradeoff factor. Finally, the overall performance loss of a specific detector (such as the ZF detector) compared to communication-only systems is the sum of the C-Loss and the MI-Loss.
					\end{remark}
					
					\begin{remark}
						The optimal tradeoff factor of the ZF detector is a function of both $P_{\rm{s}}$ and $\sigma^2$. This indicates that, when a specific MIMO signal detection algorithm is applied, the optimal tradeoff factor that achieves the best signal detection performance is related to the S\&C power and noise density. In other words, the optimal tradeoff factor should be environment-adaptive, and we cannot expect one FP-type receiver, such as the projection-type receiver or the SIC-type receiver, to achieve the desired performance in various scenarios.
					\end{remark}}
					
{The previous analysis reveals that receiver performance heavily depends on the choice of detector. For instance, projection-type receivers favor powerful detectors like the ML detector, while conventional SIC is better suited for simpler MIMO detectors like the ZF detector. Most importantly, these results demonstrate that the subsequent algorithm design must account for the impact of the tradeoff factor $\rho$. Thus, developing a suitable detection algorithm for Problem (17) requires jointly optimizing the signal detector choice and $\rho$. This insight is applied to the low-complexity algorithm design in the subsequent section.}

				

				\section{Low-Complexity Algorithm Development}
				Deriving a suitable algorithm for solving the transformed signal detection problem in (\ref{fbgfbg}) is quite challenging due to its large dimension and ill-conditioned nature. Specifically, most low-complexity algorithms, such as message passing \cite{Khani2020Adaptive,7501561} and approximate matrix inversion \cite{9790338}, do not exploit the alphabet constraint in algorithm development, yielding poor performance when the equivalent channel matrix is ill-conditioned or even rank-deficient. On the other hand, many performance-guaranteed algorithms in rank-deficient environments, such as sphere decoding \cite{wong2007efficient} and the semidefinite relaxation (SDR) method \cite{SDRMa}, often entail high computational complexity considering the high dimensionality of $\mathbf{G}$.

				In this section, we first introduce the homotopy optimization framework, which is then used to solve the signal detection problem in (\ref{fbgfbg}). {Finally, a tailored low-complexity solution, namely, the DFP-type receiver, is presented}. As compared to the FP-type receiver, the DFP-type receiver can achieve better path-tracing performance and enhanced environmental adaptability.
				
				\subsection{Homotopy Optimization Framework}
				In general, the homotopy optimization method is a type of path-tracing algorithm that is designed to solve complex, nonlinear, and often non-convex optimization problems.  The central idea of homotopy optimization is to transform the original difficult optimization problem into a set of simpler problems by introducing a continuous transformation parameter. This parameter gradually deforms the objective function from a simple starting function to the target function that represents the original problem\cite{dunlavy2005homotopy}.

				The key to homotopy optimization is to find a proper surrogate function called the homotopy function. Assuming that we are interested in finding the optimal solution to the following problem via the homotopy optimization method
				\begin{equation}
					\mathbf{z}^{*}=\mathop {\operatorname{argmin} }f(\mathbf{z}). 
				\end{equation}
				We need to find a continuous homotopy function $h(\mathbf{z},\eta)$ so that
				\[
				h(\mathbf{z}, \eta) =
				\begin{cases}
					f^0(\mathbf{z}), & \text{if } \eta  = 0 \text{ and} \\
					\cdots \\
					f^1(\mathbf{z}), & \text{if } \eta  = 1,
				\end{cases}
				\]
				where $	f^0(\mathbf{z})=f(\mathbf{z})$. It is much easier to find the minimizer of $f^1(\mathbf{z})$  than that of $	f^0(\mathbf{z})$. 
				
				Homotopy optimization starts by solving the minimization problem with $f^1(\mathbf{z})$ as the objective function. By gradually adjusting $\eta$ from 1 to 0, the problem evolves, and the solution of the previous minimization problem serves as an initial point of the newly formulated problem. This allows the optimizer to trace the path of the solution, and as the transformation progresses, the minimizer converges to the minimizer of the original challenging problem.
				Overall, the implementation details of the homotopy optimization method are summarized in Algorithm 1.
				\begin{algorithm}
					\caption{Homotopy Optimization Framework} 
					\begin{algorithmic}	
						\STATE \textbf{Input}: Initial point $\mathbf{z}^{0}$, a decreasing sequence $\eta^{l}, l=1,\cdots, l_{\rm{max}}$ that satisfies $\eta^1=1$, $\eta^{l_{\rm{max}}}=0$.
						\STATE \textbf{Output}: Optimized result $\tilde{\mathbf{z}}={\mathbf{z}}^{l_{\rm{max}}}$.
					\end{algorithmic}	
					\begin{algorithmic}[1]
						\STATE $l=1$
						\STATE  \textbf{Repeat} 
						\STATE Use a local minimization method, starting with ${\mathbf{z}}^{(l-1)}$, to compute 
						an (approximate) solution ${\mathbf{z}}^{(l)}$ to $\mathop {\operatorname{argmin} } h(\mathbf{z}, \eta^{l})$. 
						\STATE $l=l+1$.
						\STATE  \textbf{Until $l={l_{\rm{max}}}$}	
					\end{algorithmic}
				\end{algorithm}

				
				\subsection{Homotopy-based MIMO Signal Detection Algorithm}
				{In this subsection, we introduce a homotopy-based MIMO signal detection algorithm presented in \cite{9311778, 10677490}. The authors of \cite{9311778, 10677490} first found the homotopy function by proposing an equivalent transformation of the MIMO signal detection problem. Subsequently, a projected gradient (PG) method is used under the homotopy optimization framework. In the following, we explain how to use this MIMO signal detection algorithm to solve Problem (\ref{fbgfbg}) in the FP-type receiver, which also serves as the basis of our subsequent design.  }

				By letting 
				\vspace{-0.1cm}
				\begin{equation} \label{nngfgh}
					\overline{\mathbf{y}}=\left[\begin{array}{l}
						\operatorname{Re} (\mathbf{y}_{\rm{FP}}) \\
						\operatorname{Im} ({\mathbf{y}_{\rm{FP}}})
					\end{array}\right], \quad \overline{\mathbf{x}}=\left[\begin{array}{c}
						\operatorname{Re}({\mathbf{x}}_{\mathrm{c}})\\
						\operatorname{Im}({\mathbf{x}}_{\mathrm{c}})
					\end{array}\right],
				\end{equation}
				and
				\begin{equation} \label{nngfghv}
					\overline{\mathbf{G}}=\left[\begin{array}{cc}
						\operatorname{Re}(\mathbf{G}_{\rm{FP}}) & -\operatorname{Im}(\mathbf{G}_{\rm{FP}}) \\
						\operatorname{Im}(\mathbf{G}_{\rm{FP}}) & 	\operatorname{Re}(\mathbf{G}_{\rm{FP}})
					\end{array}\right],
				\end{equation} 
				Problem (\ref{fbgfbg})   can be rewritten as
				\begin{equation} \label{fbgf}
					\min _{\overline{\mathbf{x}}\in  \mathcal{X}_{\mathbb{R}}^{2LK}} f(\overline{\mathbf{x}} )=\|\overline{\mathbf{y}} -\overline{\mathbf{G}}  \overline{\mathbf{x}} \|_{2}^{2},
				\end{equation}
{where $\overline{\mathbf{x}}$ is a $2LK$-dimensional real-valued vector whose entries lie in the finite real-valued alphabet set $\mathcal{X}_{\mathbb{R}}$.}
				
				The main challenge of solving Problem (\ref{fbgf}) lies in the discrete alphabet constraint. In the following theorem, we show that the homotopy function can be constructed by relaxing the alphabet constraint into its convex hull and penalization.

				\begin{theorem}
					Problem (\ref{cs}) is an equivalent transformation of Problem (\ref{fbgf}) if the penalty parameter $\mu> \lambda_{1}(\overline{\mathbf{G}}^{\rm{T}} \overline{\mathbf{G}})$,	
				\begin{equation} \label{cs}
					\begin{aligned}
						\min _{\overline{\mathbf{x}} \in \text{conv}(\mathcal{X}_{\mathbb{R}}^{2LK})} &F_{\mu}(\overline{\mathbf{x}})=f(\overline{\mathbf{x}})-\mu\|\overline{\mathbf{x}}\|_{2}^{2}\\
						&=\overline{\mathbf{x}}^{\rm{T}}\left(\overline{\mathbf{G}}^{\rm{T}} \overline{\mathbf{G}}-\mu \mathbf{I}\right) \overline{\mathbf{x}}-2 \overline{\mathbf{y}}^{\rm{T}} \overline{\mathbf{G}}^{\rm{T}} \overline{\mathbf{x}}+\|\overline{\mathbf{y}}\|_{2}^{2}.
					\end{aligned}
				\end{equation}
				\end{theorem}
				
				\begin{proof}
					$F_{\mu}(\overline{\mathbf{x}})$ is strictly concave since  $\overline{\mathbf{G}}^{\rm{T}} \overline{\mathbf{G}}-\mu \mathbf{I}$ is negative definite. Theorem 2 can be proved by using the conclusion in \cite[Section 32]{rockafellar1970convex}.
				\end{proof}
				
				
				
				While Problem (\ref{cs}) is still a highly non-convex and hard-to-solve problem due to its non-concave objective function, the homotopy optimization method can be adopted with a gradually increasing penalty sequence $\mu_l, l=1,\cdots, l_{\rm{max}}$.  Intuitively, when the penalty parameter $\mu$ is small, the gap between Problem (\ref{fbgf}) and Problem (\ref{cs}) is large, which potentially means that the constraint ${\overline{\mathbf{x}} \in \mathcal{X}_{\rm{R}}}$ may not be satisfied. On the other hand, the convexity of Problem (\ref{cs}) is better preserved, suggesting that it is easier to obtain the optimal solution of Problem (\ref{cs}).
				Considering these properties, we start with a small value of $\mu$ (i.e., $\mu_1=0$) and solve Problem (\ref{cs}) with gradually increasing $\mu_l$ in the $l$-th iteration. Finally, with a sufficiently large value of $\mu_l$, we can gradually approach the optimal solution to Problem (\ref{fbgfbg}).

				With a fixed value of $\mu_l$, Problem (\ref{cs}) can be solved using the PG method \cite{beck2017first}. The PG method is a low-complexity iterative algorithm that alternates between gradient descent and projection. In conventional MIMO signal detection problems, such as Problem (\ref{fbgf}), projection involves rounding to the discrete constellation, leading to conflicting effects as the performance of the PG method approaches that of the linear decoder \cite{10181137}.  Specifically, during the gradient descent stage, the solutions move toward the unconstrained LS problem, while the subsequent projection to the discrete constellation diminishes the effectiveness of the earlier process.
				
				When the original MIMO signal detection problem is rewritten into the continuous function in (\ref{cs}), the gradient descent step and the projection step of the PG algorithm have been modified. First, gradient descent is applied at the majorant of $F_{\mu}(\overline{\mathbf{x}})$. The majorant of $F_{\mu}(\overline{\mathbf{x}})$ at the $k$-th step $\overline{\mathbf{x}}^{k}$ is obtained by replacing $\|\overline{\mathbf{x}}\|_{2}^{2}$ with its first-order approximation, given by
				\begin{equation}
					G_{\mu}(\overline{\mathbf{x}} \mid \overline{\mathbf{x}}^k)=f(\overline{\mathbf{x}})-2 \mu_l \langle \overline{\mathbf{x}}^k, \overline{\mathbf{x}}-\overline{\mathbf{x}}^k\rangle-\mu_l \|\overline{\mathbf{x}}^k \|_2^{2}.
				\end{equation}
				Thus, the gradient of $G_{\mu}(\overline{\mathbf{x}} \mid \overline{\mathbf{x}}^k)$ is given by
				\begin{equation} \label{nfn}
					\begin{aligned}
						\nabla_{\overline{\mathbf{x}}} G_{\mu}\left({\overline{\mathbf{x}}} \mid {\overline{\mathbf{x}}^k}\right)&= 2\overline{\mathbf{G}}^{\rm{T}} \overline{\mathbf{G}} {\overline{\mathbf{x}}}- 2\overline{\mathbf{G}}^{\rm{T}} \overline{\mathbf{y}}-2  \mu_{l} \overline{\mathbf{x}}^k.
					\end{aligned}
				\end{equation}
				Second, the projection operation is performed on the continuous convex hull of the original alphabet constraints $\text{conv}(\mathcal{X}_{\rm{R}}^{2LK})$. Given that the alphabet constraints in most modulation schemes, such as phase shift keying (PSK) and quadrature amplitude modulation (QAM), can be written as constant-modulus constraints \cite{10677490}, the projection onto these constraints is usually computationally effective and also better at using gradient information, thereby accelerating convergence speed.
				
			{	Here we take 4-QAM as an example to show the benefit of applying the transformation in Theorem 2. When the PG method is directly applied for solving Problem (\ref{fbgf}), the projection is performed at the constellation constraint $\overline{\mathbf{x}} \in \{-1,1\}^{n}$, which means deciding each transmitted symbol according to its sign. In contrast, with the homotopy optimization, the projection is performed at the convex hull of the constellation constraint, $\operatorname{conv}\left(\{-1,1\}^{n}\right)=[-1,1]^{n}$. As a result, the projection operation becomes $ \Pi_{\mathcal{X}}(\mathbf{x})=\left[\Pi\left(x_{1}\right), \ldots, \Pi\left(x_{N}\right)\right]^{\rm{T}}$, where
				\begin{equation} \label{bgf}
					\Pi(x)=\left\{\begin{array}{ll}
						-1, & x<-1 \\
						x, & -1 \leq x \leq 1. \\
						1, & x>1
					\end{array}\right.
				\end{equation}
				Thus, the projection in (\ref{bgf}) tends to become smoother for $-1 \leq x \leq 1$, making the PG method much easier to escape from the locally optimal solution.}

				\subsection{DFP-Type Receiver Design}
				While the aforementioned algorithm can efficiently address the transformed MIMO signal detection problem, it still faces the dilemma of selecting the proper value of $\rho$. 
				Moreover, the FP-type receiver fails to establish the equivalence between Problem (\ref{fbgfbg}) and Problem (\ref{eqvvq}). In this regard, we further apply the spirit of homotopy optimization, which leads to the DFP-type receiver. In Lemma 5, we provide a theoretical foundation for the proposed DFP-type receiver. 
				
				\begin{lemma}
					Problem (\ref{cst}) is an equivalent transformation of Problem (\ref{vbdf}) if $\mu> \lambda_{1}(\mathbf{H}_{\rm{c}}^{\rm{H}}\mathbf{H}_{\rm{c}})$, and $\rho=0$,
				\begin{equation} \label{cst}
					\begin{aligned}
						\min _{\overline{\mathbf{x}} \in \text{conv}(\mathcal{X}_{\mathbb{R}}^{2LK})} F_{\mu, \rho}(\overline{\mathbf{x}})&=\overline{\mathbf{x}}^{\rm{T}}\left(\overline{\mathbf{G}}^{\rm{T}}(\rho) \overline{\mathbf{G}}(\rho)-\mu \mathbf{I}\right) \overline{\mathbf{x}}\\
						&-2 \overline{\mathbf{y}}^{\rm{T}}(\rho) \overline{\mathbf{G}}^{\rm{T}}(\rho) \overline{\mathbf{x}}+\|\overline{\mathbf{y}}\|_{2}^{2},
					\end{aligned}
				\end{equation}
				where  $\overline{\mathbf{y}}(\rho)$ and $\overline{\mathbf{G}}(\rho)$ are a function of $\rho$ as defined in (\ref{nngfgh}) and (\ref{nngfghv}), respectively.
				\end{lemma}
				
				\begin{proof}
					Please refer to Appendix C.
				\end{proof}
				
				\begin{remark}
					Different from the conventional homotopy optimization framework, which introduces an auxiliary variable to transform complex problems into more solvable ones, two auxiliary variables are introduced to establish the equivalence between the newly formulated Problem (\ref{cst}) and the original problem in (\ref{vbdf}). Therefore, more DoFs in algorithm development are provided.  In the following, we introduce the structure and implementation details of the DFP-type receiver.
				\end{remark}
				
				As shown in Fig. 6, the communication signal detection process in the DFP-type receiver can be interpreted as a two-loop iterative algorithm that can be divided into three parts. 
				
				In the $l$-th outer iteration,  both the penalty parameter $\mu_{l}$ and the tradeoff factor $\rho_{l}$  are updated, as shown in the first part of {Fig. 6}. The first part is the major difference between the DFP-type receiver and the FP-type receiver. 
				Particularly, the tradeoff factor $\rho$ is updated in each outer-layer iteration according to the predefined sequence $\rho_l, l=0,\cdots, l_{\textrm{max}}$, and $\rho_l$ at the $l$-th iteration is heuristically chosen as   
				\begin{equation}
					\rho_l=\epsilon^{l},  0<\epsilon<1.
				\end{equation} 
				This exponential expression ensures that $\rho_0 =1$ and, with an adequate number of iterations, $\rho_l \to 0$. Then, in each iteration, the matrix $\mathbf{P}_{l}=\mathbf{P}(\rho_l)$, the vector $\mathbf{y}_{l}=\overline{\mathbf{y}}(\rho_l) $, and the matrix $\mathbf{G}_{l}=\overline{\mathbf{G}}(\rho_l)$ are updated according to (\ref{bbdf}), (\ref{ngf}) together with (\ref{nngfgh}), and (\ref{ngf}) together with (\ref{nngfghv}), respectively. 
				
				\begin{figure} \label{bdbb}
					\centering	
					\includegraphics[width=9cm]{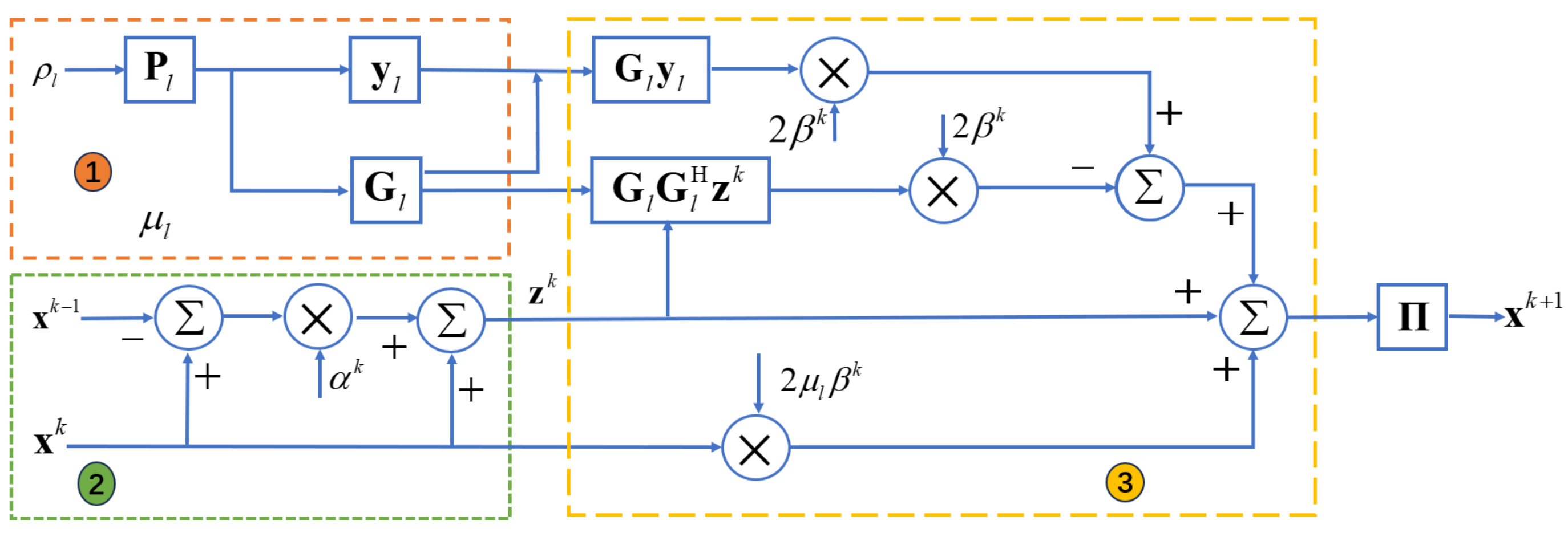}
					\caption{\small{Illustration of the proposed DFP-type receiver design}}
					
				\end{figure}

				In the inner layer, the PG algorithm is applied, as shown in the second and the third parts of {Fig. 6}. The second part includes a modification of the conventional gradient descent algorithm, namely gradient-extrapolated majorization-minimization \cite{8811616}. More specifically, the gradient descent at the $k$-th step is performed on the linear combination of the current point $\overline{\mathbf{x}}^k$ and the previous point $\overline{\mathbf{x}}^{k-1}$, which is called the extrapolated point $\mathbf{z}^{k}$ as follows
				\begin{equation} \label{bet}
					\mathbf{z}^{k}=\alpha_k(\overline{\mathbf{x}}^k-\overline{\mathbf{x}}^{k-1})+\overline{\mathbf{x}}^k.
				\end{equation}
				Here, the linear combination coefficients sequence {$\alpha_{k}$} is chosen from a predefined sequence as the design in \cite{9311778}.

				The third part is to perform gradient descent on the extrapolated point $\mathbf{z}^{k}$. The gradient descent step size $\beta^{k}$ is chosen as $\beta^{k}=1/\|\mathbf{G}_l\|_2$ to fulfill the sufficient descent condition of the quadratic function. Then, the point after gradient descent $\mathbf{t}^{k}$ is given by
				\begin{equation} \label{nfnv}
					\begin{aligned}
						\mathbf{t}^{k}&=\mathbf{z}^{k}-{\beta^{k}} \nabla_{\overline{\mathbf{x}}} G_{\mu}\left(\mathbf{z}^{k} \mid {\overline{\mathbf{x}}^k}\right) \\
						&=\mathbf{z}^{k}-2\beta^{k}{\mathbf{G}}_l^{\rm{T}} {\mathbf{G}}_l \mathbf{z}^{k}+2\beta^{k} {\mathbf{G}}_l^{\rm{T}} {\mathbf{y}}_l+2 \beta^{k} \mu_{l} \overline{\mathbf{x}}^k.
					\end{aligned}
				\end{equation}
		{In the conventional FP algorithm, a fixed tradeoff parameter $\rho$ is adopted. Therefore, the high-dimensional matrix products required for the gradient computation, such as $\mathbf{G}^{\rm{T}}\mathbf{G}$ and $\mathbf{G}^{\rm{T}}\mathbf{y}$, remain constant and only need to be calculated once. However, a major computational bottleneck in the proposed DFP algorithm lies in the dynamic nature of $\rho_l$. Since $\rho_l$ is updated at each outer iteration $l$, directly recomputing the real-valued matrix multiplications $\mathbf{G}_l^{\rm{T}} \mathbf{G}_l$ and $\mathbf{G}_l^{\rm{T}} \mathbf{y}_l$ at every iteration would incur a prohibitive computational burden. To address this issue, we exploit the linear nature of the real-valued decomposition defined in (\ref{nngfgh}) and (\ref{nngfghv}).
			
			Specifically, in the complex domain, the matrix $\mathbf{G}_{\rm{FP}}$ and the vector $\mathbf{y}_{\rm{FP}}$ are formulated as a linear combination weighted by $(1-\rho_l)$ and $\rho_l$. Because extracting the real and imaginary parts is a strictly linear operation, this combination structure is perfectly preserved in the real-valued domain. Consequently, the real-valued matrix $\mathbf{G}_l$ and vector $\mathbf{y}_l$ at the $l$-th iteration can be equivalently decomposed as
			\begin{equation} \label{real_affine_G}
				\mathbf{G}_l = (1-\rho_l){\mathbf{G}}_0 + \rho_l{\mathbf{G}}_1,
			\end{equation}
			\begin{equation} \label{real_affine_y}
				\mathbf{y}_l = (1-\rho_l){\mathbf{y}}_0 + \rho_l{\mathbf{y}}_1,
			\end{equation}
			where ${\mathbf{G}}_0, {\mathbf{G}}_1$ and ${\mathbf{y}}_0, {\mathbf{y}}_1$  are completely independent of the iteration index $l$ and the parameter $\rho_l$.
			
			Substituting (\ref{real_affine_G}) and (\ref{real_affine_y}) into the gradient terms, we can reformulate the iteration-dependent matrix products into polynomials of $\rho_l$
			\begin{equation} \label{GTG_real_fast}
				\mathbf{G}_l^{\rm{T}} \mathbf{G}_l = (1-\rho_l)^2 \mathbf{M}_1 + \rho_l(1-\rho_l) \mathbf{M}_2 + \rho_l^2 \mathbf{M}_3,
			\end{equation}
			\begin{equation} \label{Gy_real_fast}
				\mathbf{G}_l^{\rm{T}} \mathbf{y}_l = (1-\rho_l)^2 \mathbf{u}_1 + \rho_l(1-\rho_l) \mathbf{u}_2 + \rho_l^2 \mathbf{u}_3,
			\end{equation}
			where the constant matrices are given by 
			\begin{equation} \label{berer}
\mathbf{M}_1 = \overline{\mathbf{G}}_0^{\rm{T}} \overline{\mathbf{G}}_0, \; \mathbf{M}_2 = \overline{\mathbf{G}}_0^{\rm{T}} \overline{\mathbf{G}}_1 + \overline{\mathbf{G}}_1^{\rm{T}} \overline{\mathbf{G}}_0, \; \mathbf{M}_3 = \overline{\mathbf{G}}_1^{\rm{T}} \overline{\mathbf{G}}_1,
			\end{equation}
	 and the constant vectors are given by
	 \begin{equation} \label{betb}
			\mathbf{u}_1 = \overline{\mathbf{G}}_0^{\rm{T}} \overline{\mathbf{y}}_0,\;\mathbf{u}_2 = \overline{\mathbf{G}}_0^{\rm{T}} \overline{\mathbf{y}}_1 + \overline{\mathbf{G}}_1^{\rm{T}} \overline{\mathbf{y}}_0, \; \mathbf{u}_3 = \overline{\mathbf{G}}_1^{\rm{T}} \overline{\mathbf{y}}_1.
	 \end{equation}
			By explicitly pre-computing the constant matrices $\{\mathbf{M}_i\}_{i=1}^3$ and vectors $\{\mathbf{u}_i\}_{i=1}^3$ before the iterations start, the gradient computation at each step is simplified to basic scalar multiplications and matrix additions. Through this structural design, the proposed DFP algorithm bypasses the repeated calculation of large-scale matrix multiplications entirely, thereby achieving the exact same order of computational complexity as the conventional FP algorithm.}
				
				Finally, the estimated transmit signal at the $(k+1)$-th iteration can be obtained by projecting $\mathbf{t}^{k}$ onto the convex hull of the real-valued alphabet constraint $\text{conv}(\mathcal{X}_{\rm{R}})$, 
				\begin{equation}\label{nf}
					\overline{\mathbf{x}}^{k+1}= \Pi\left(\mathbf{t}^{k}\right).
				\end{equation}
				
				Since gradient descent is adopted, receiver performance is affected by the initial point. Therefore, it is crucial to select an initial point that not only enhances algorithm performance but also maintains low computational complexity. In the DFP-type receiver, we start with the ZF decoder. By exploiting the Kronecker structure, the inverse in the ZF detector can be computed efficiently: $\mathbf{G}_{\rm{FP}}^{\dagger}=\mathbf{P}_{\rm{FP}}^{\dagger} \otimes \mathbf{H}_{\rm{c}}^{\dagger}$. Finally, the overall algorithm of the DFP-type receiver is summarized in Algorithm 2.

\begin{algorithm}
	\caption{Proposed Algorithm for the DFP-Type Receiver}
	\begin{algorithmic} [1]
		\STATE  \textbf{Input}: The maximum number of outer iterations $l_{\textrm{max}}$, the maximum number of inner iterations $k_{\textrm{max}}$, the penalty sequence $\{\mu_l\}$, the tradeoff sequence $\{\rho_l\}$, the step size sequence $\{\beta^k\}$ and an initial point $\overline{\mathbf{x}}_0$.	
		\STATE  \textbf{Output}: Estimated communication signal $\hat{{\mathbf{x}}}_{\mathrm{c}}$, and estimated target response vector $\hat{\mathbf{h}}_{\rm{r}}$.
		
		
		\REPEAT 
		
		
		\REPEAT
		\STATE Update the extrapolated point $\mathbf{z}^{k}$ according to (\ref{bet}).
		\STATE Obtain $\mathbf{t}^{k}$ by using the gradient descent in (\ref{nfnv}).
		\STATE Obtain $\overline{\mathbf{x}}^{k+1}$ by projecting $\mathbf{t}^{k}$ using (\ref{nf}).
		
		{\IF {$F_{\mu_l, \rho_l}(\overline{\mathbf{x}}^{k+1}) < f_{\min}$}
		\STATE $f_{\min} \gets F_{\mu_l, \rho_l}(\overline{\mathbf{x}}^{k+1})$
		\STATE $\mathbf{x}_{\textrm{best}} \gets \overline{\mathbf{x}}^{k+1}$
		\ENDIF }
		
		\STATE $k\gets k+1$.
		\UNTIL $k=k_{\textrm{max}}.$
		
		\STATE  $l \gets l + 1$
		\UNTIL  $l=l_{\textrm{max}}.$
		
		\STATE  $\hat{{\mathbf{x}}}_{\mathrm{c}}=\overline{\mathbf{x}}_{l}.$
		\STATE Estimate the target response vector $\hat{\mathbf{h}}_{\rm{r}}$ with the decoded signal $\hat{{\mathbf{x}}}_{\mathrm{c}}$ according to (\ref{bav}).
	\end{algorithmic}
\end{algorithm}
				
				
				In summary, adopting the DFP-type receiver instead of the fixed-tradeoff-factor FP-type receiver has two benefits. First, the proposed algorithm provides a smoother path for tracing the solution of the original problem, thereby reducing the difficulty of solving Problem (\ref{vbdf}) or, equivalently, Problem (\ref{eqvvq}). Second, although we cannot obtain the ``optimal and environment-adaptive'' tradeoff factor discussed in Remark 5, we use a heuristic search with greater diversity than that of the fixed-tradeoff-factor FP-type receiver. Hence, the DFP-type receiver is expected to be more robust in various scenarios.

				\subsection{Complexity Analysis}	
				To precisely evaluate the computational efficiency of the proposed DFP receiver and the conventional FP receiver, we quantify the exact number of floating-point operations (FLOPs) for the standard multiplication or addition in the real-valued domain. For the DFP and FP algorithms, which perform joint detection over the entire snapshot sequence, the optimization variable $\overline{\mathbf{x}}$ is a $2LK \times 1$ vector. During the iterative process, the dominant computational burden lies in the inner-loop gradient descent step, specifically the matrix-vector multiplication between the high-dimensional real-valued gradient matrix $\mathbf{G}_l^{\rm{T}} \mathbf{G}_l \in \mathbb{R}^{2LK \times 2LK}$ and the extrapolated vector $\mathbf{z}^k \in \mathbb{R}^{2LK \times 1}$. This full-scale multiplication requires exactly $(2LK)^2$ multiplications and $2LK(2LK-1)$ additions, yielding approximately $8L^2K^2$ FLOPs per inner iteration. In summary, the explicit total computational complexity for the proposed DFP receiver is formulated as
				\begin{align} \label{flops_dfp}
				\text{FLOPs (DFP)} =& C_{\text{ZF}}K^3 + l_{\max} \times \\ &\left( 12L^2K^2 + k_{\max} \times 8L^2K^2 \right), \nonumber
				\end{align}
				where $C_{\text{ZF}}K^3$ represents the initialization overhead of the ZF decoder, and $l_{\max}$ and $k_{\max}$ denote the maximum number of outer and inner iterations, respectively.
				
				As a special class of the FP-type receivers, the SIC algorithm exhibits a significantly lower exact computational complexity, fundamentally because it bypasses the joint processing of multiple symbols across the $L$ snapshots. Specifically, in the SIC-based receiver, the equivalent projection matrix is structurally fixed as $\mathbf{G}_{\text{SIC}} = \mathbf{I}_L \otimes \mathbf{H}_{\rm{c}}$. By exploiting the mixed-product property of the Kronecker product, the core high-dimensional matrix required for the gradient computation is  decoupled as
				\begin{equation} \label{kron_sic}
				\mathbf{G}_{\text{SIC}}^{\rm{H}} \mathbf{G}_{\text{SIC}} = (\mathbf{I}_L \otimes \mathbf{H}_{\rm{c}})^{\rm{H}} (\mathbf{I}_L \otimes \mathbf{H}_{\rm{c}}) = \mathbf{I}_L \otimes (\mathbf{H}_{\rm{c}}^{ \rm{H}} \mathbf{H}_{\rm{c}}).
				\end{equation}
				This indicates that the above $2LK \times 2LK$ matrix perfectly reduces to a block-diagonal structure comprising $L$ independent $2K \times 2K$ blocks in the real-valued domain. Consequently, the large-scale matrix-vector multiplication is completely decomposed into $L$ parallel and independent sub-multiplications. Since each $2K \times 2K$ block multiplication strictly requires $8K^2$ FLOPs, the total computational cost per iteration for the SIC receiver is drastically reduced to $8LK^2$ FLOPs. Accordingly, the overall computational complexity for the baseline SIC receiver can be explicitly given by	
				\begin{align} \label{flops_sic}
					\text{FLOPs (SIC)} =& C_{\text{ZF}}K^3 + l_{\max} \times k_{\max} \times 8LK^2.
				\end{align}
				The FLOPs analysis reveals that the proposed joint detection incurs a complexity precisely $L$ times that of the SIC scheme per iteration, representing a necessary tradeoff to unlock the substantial performance gains inherent in joint temporal processing.

					\subsection{PDFP-Type Receiver Design}
					
					The DFP-type receiver significantly enhances environment adaptation compared to the FP-type receiver. However, its adaptive ability is still limited by the selection of the predefined tradeoff sequence  $\rho_l=\epsilon^l$. { For instance, when $\epsilon$ is small, the receiver quickly converges to the projection-type receiver due to the exponential sequence. While this completely eliminates sensing interference, it inherently results in an underdetermined observation matrix. Thus, this design is more suitable for scenarios with strong sensing signals, but less effective for weak ones where the penalty of an underdetermined system outweighs the benefits of interference nulling.} To further improve environment adaptation, a parallel dynamic flexible projection (PDFP)-type receiver can be used, consisting of multiple DFP-type receivers.  In the PDFP-type receiver, we assume that $P$ DFP-type receivers simultaneously perform communication signal detection using different convergence speeds. The decreasing sequence of the tradeoff factor for the $p$-th receiver is chosen as $\rho_{l,p}=\epsilon_p^{l}, p=1,\cdots, P$. Once the estimated communication signal $\hat{\mathbf{x}}_{{\rm{c}},p} $ and the target response vector $\hat{\mathbf{h}}_{{\rm{r}},p}$ of all $P$ receivers are obtained, the final communication signal and target response matrix are chosen based on the detector with the minimum square error:
					\begin{equation}
						\hat{\mathbf{x}}_{\rm{c}}, \hat{\mathbf{h}}_{\rm{r}} = \mathop {\arg \min}\limits_{\hat{\mathbf{x}}_{{\rm{c}},p},\hat{\mathbf{h}}_{{\rm{r}},p}, p\in \{1,...,P\}} \left\| \mathbf{y} - \mathbf{A}_{\rm{c}}\hat{\mathbf{x}}_{{\rm{c}},p} - \mathbf{A}_{\rm{r}}\hat{\mathbf{h}}_{{\rm{r}},p} \right\|_2^2.
					\end{equation}

								\section{Simulation Results}
								Simulation results are provided to evaluate the performance of the uplink ISAC systems. We assume 4-QAM modulation for uplink communication and orthogonal radar waveforms transmitted from different antennas, i.e., $\mathbf{R}=P_{ \rm{r}}/{M_t}\mathbf{I}$. Unless stated otherwise, the simulation parameters are set as follows. The numbers of transmit antennas, receive antennas, and CUs are set to $M_t=4$, $M_r=8$, and $K=8$, respectively. We assume that the communication channel satisfies $\operatorname{vec}(\mathbf{H}_{\rm{c}}) \sim{\mathcal C}{\mathcal N}\left( {{\mathbf{0}}, {{\mathbf{I}}_{KM_r}}} \right) $ and that $L=16$ snapshots are jointly processed. The normalized transmit power of the communication signal is set to $P_{\rm{c}}=1$ W, while the noise power is set to $\sigma^2=-10$ dBW\footnote{The noise power is normalized by the pathloss and the actual thermal noise at the receiver.}. The penalty parameter is initialized as $\mu_{0}=0.001$, with a standard step size in the subgradient method and the update rule presented in \cite{10677490}. The tradeoff factor in each iteration is updated as $\rho_l=0.05^{l}$. The maximum numbers of outer and inner iterations are set as $l_{\rm{max}}=200$ and $k_{\rm{max}}=100$, respectively. We use the homotopy optimization method outlined in Subsection V-B for the SIC-type, projection-type, and FP-type receivers, with parameters set to the same values as those for the DFP-type receiver.


								\subsection{Performance Evaluation of the Proposed Receiver}
								
									\begin{figure*}
									\vspace{-1cm}
									\centering
									\begin{minipage}[t]{0.49\textwidth}
										\centering
										\includegraphics[width=8cm]{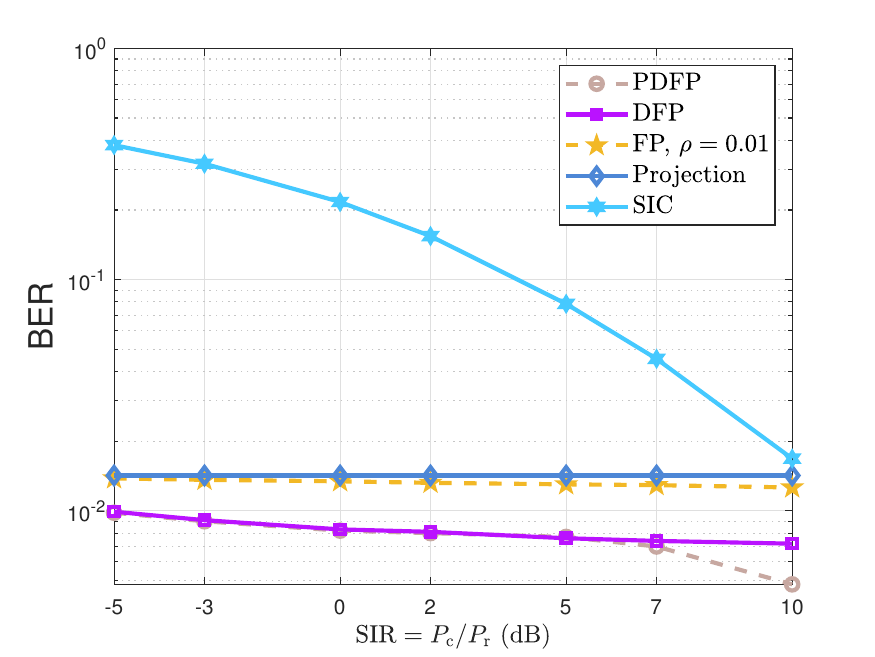}
										\caption{\small{BER of different types of receivers}}
									\end{minipage}
									\begin{minipage}[t]{0.49\textwidth}
										\centering
										\includegraphics[width=8cm]{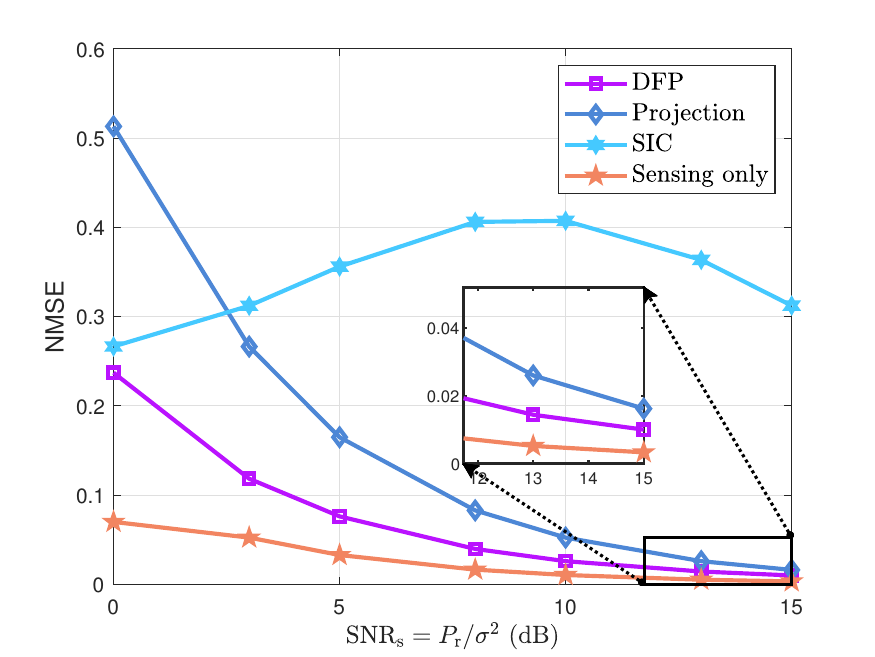}
										\caption{\small{Normalized mean-square error (NMSE) versus the sensing SNR}}
									\end{minipage}
								\end{figure*}
In this subsection, we evaluate the S\&C performance tradeoff under different S\&C power levels, as shown in Figs. 7 and 8. In Fig. 7, we examine the BER of different receivers under different signal-to-interference ratios (SIRs) for the communication signal detection problem, where $\textrm{SIR}\triangleq P_{\rm{c}}/P_{\rm{r}} $. As observed in Fig. 7, the projection-type receiver exhibits a constant BER across different SIR levels, indicating a lack of environmental adaptation. The performance of the SIC-type receiver deteriorates when the SIR is below 15 dB, suggesting that it is effective only when the sensing signal is weak. In contrast, the proposed DFP-type receiver outperforms both the projection-type and FP-type receivers, especially at higher SIR levels. However, its performance begins to saturate when the SIR exceeds 10 dB. The PDFP-type receiver, which combines two DFP-type receivers to enhance environmental adaptation, provides improved performance across a wider range of SIR values.

In Fig. 8, we examine the NMSE of the target response vector ${\mathbf{h}}_{\rm{r}}$ achieved by different receivers as the transmit power of the sensing signal varies. The sensing SNR, corresponding to the target response estimation problem, is defined as $\textrm{SNR}_{\rm{s}}\triangleq P_{\rm{r}}/ \sigma^2$. It can be observed that the NMSEs of the projection-type and DFP-type receivers decrease as the transmit power increases and become lower than 0.02 when $\textrm{SNR}_{\rm{s}}=15$ dB. Moreover, the DFP-type receiver outperforms the FP-type receiver because it addresses the communication signal detection problem more efficiently. However, the performance of the SIC-type receiver does not improve monotonically as the sensing SNR increases. This is because, in the SIC-type receiver, communication signal detection in the first stage is poor when the transmit power of the sensing signal is high. In this case, the residual communication signal detection error is much larger than the noise, which degrades the subsequent sensing performance. Therefore, the projection-type receiver and the proposed DFP-type receiver are better suited to achieving the desired S\&C performance simultaneously.
								
Apart from the NMSE of target response estimation, the angle estimation performance of different receivers is compared in Fig. 9(a)--(c) under a bistatic sensing setup involving both AoA and AoD estimation. Given the limited numbers of transmit and receive antennas, standard FFT beamforming yields only coarse angular resolution, which necessitates the use of high-resolution algorithms. Specifically, after acquiring the target response matrix, we employ the CS-based orthogonal matching pursuit (OMP) method to estimate the AoA and AoD for multiple targets. The results indicate that the DFP-based receivers successfully resolve two distinct targets, whereas the SIC-based receiver fails to do so. Furthermore, we evaluate the angle estimation performance in terms of root mean-square error (RMSE) and hit rate, defined as the proportion of estimates with an RMSE below $2^\circ$, for different receivers, as summarized in Table I.	
	
\begin{figure*}
	\centering
	\begin{minipage}[t]{0.32\textwidth}
		\centering
		\includegraphics[width=\linewidth]{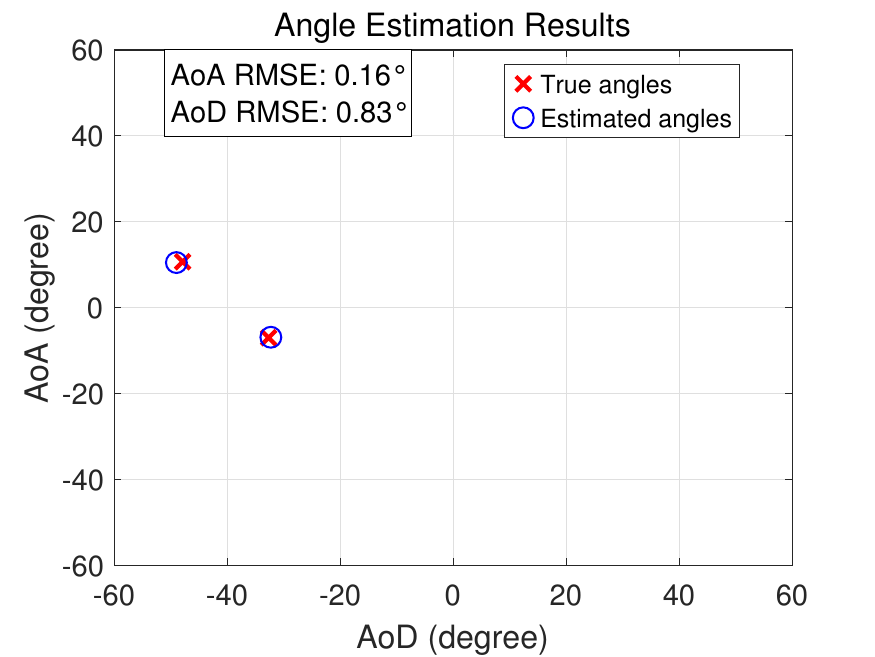}
		\caption*{\small{Fig. 9(a): Sensing-only, $M_t=M_r=K=8$}}
	\end{minipage}
	\hfill
	\begin{minipage}[t]{0.32\textwidth}
		\centering
		\includegraphics[width=\linewidth]{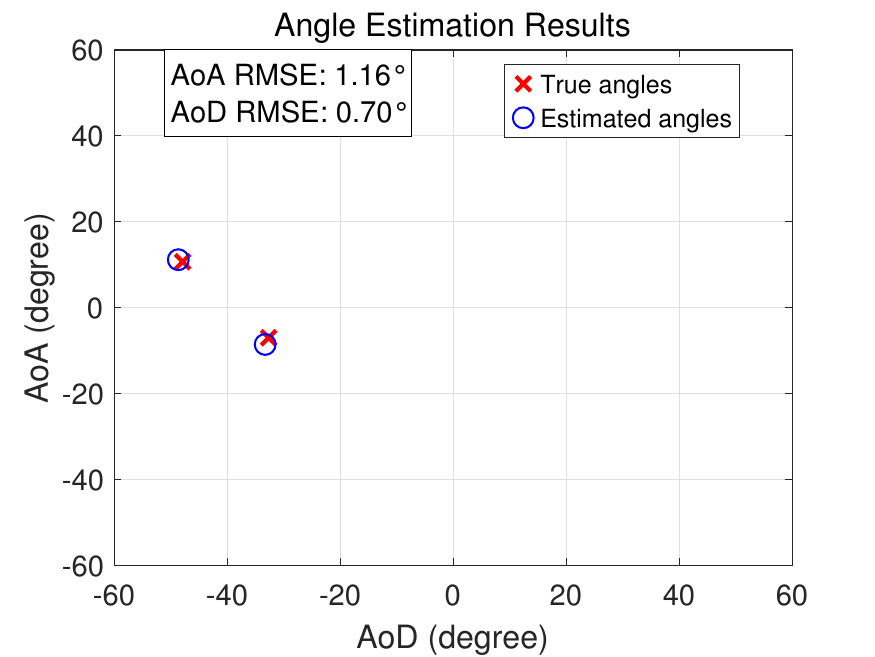}
		\caption*{\small{Fig. 9(b): DFP, $M_t=M_r=K=8$}}
	\end{minipage}
	\hfill
	\begin{minipage}[t]{0.32\textwidth}
		\centering
		\includegraphics[width=\linewidth]{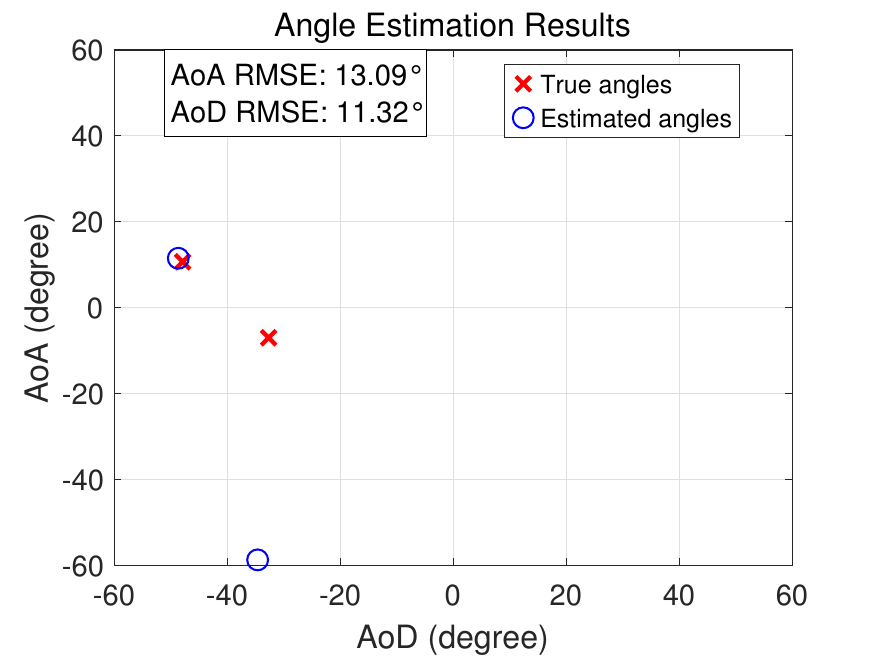}
		\caption*{\small{Fig. 9(c): SIC, $M_t=M_r=K=8$}}
	\end{minipage}
\end{figure*}			

								\begin{table}[htbp]
										\centering
										\caption{RMSE Comparison of Angle Estimation for Different Receivers}
										\label{tab:rmse_comparison}
										\renewcommand{\arraystretch}{1.3}
										\begin{tabular}{|c|c|c|c|}
											\hline
											\textbf{} & \textbf{DFP-type} & \textbf{SIC-type} & \textbf{Sensing-only} \\
											\hline
											RMSE & 5.6$^\circ$ &    8.2$^\circ$ &     2.2$^\circ$ \\
											\hline
											Hit rate & 71.3\% &    50.9\% & 90.2\% \\
											\hline
										\end{tabular}
								\end{table}
								\subsection{Impact of System Parameters}
								In Fig. 10, we evaluate the performance of different receivers under different noise power levels. The SNR of the signal detection problem is defined as $\textrm{SNR}_{\rm{c}}\triangleq P_{\rm{c}}/\sigma^2$. In Fig. 10(a), we consider a typical i.i.d. Gaussian channel, while in Fig. 10(b), we consider a correlated communication channel with a correlation coefficient of $r = 0.3$ \cite{951380}. Additionally, we examine the impact of imperfect CSI, where the estimated channel $\hat{\mathbf{H}}_{\rm{c}} = \mathbf{H}_{\rm{c}} + \Delta \mathbf{H}$ is available at the receiver. The CE error $\Delta  {\mathbf{H}}$ follows the distribution $\Delta \mathbf{H} \sim \mathcal{CN} \left( \mathbf{0}, \sigma_{\rm{e}}^2 \mathbf{I}_{M_r} \right)$ with $\sigma_{\rm{e}}^2=0.01$. It can be observed that the DFP-type receiver yields better performance than the projection-type receiver using homotopy optimization and approaches the performance of the projection-type receiver using the SDR detector. Moreover, performance degrades significantly in the presence of CE errors, underscoring the importance of accurate CE. Finally, although the performance of the uplink ISAC receiver deteriorates under correlated communication channels, the overall performance trend remains similar.
									\begin{table}[htbp]
										\centering
										\caption{Computational time of different algorithms per snapshot (s)}
										\label{tab:time_comparison}
										\renewcommand{\arraystretch}{1.3} 
										\begin{tabular}{|c|c|c|c|}
											\hline
											\textbf{$(M_r, K, L)$} & \textbf{DFP receiver} & \textbf{SIC receiver} & \textbf{Projection-SDR } \\
											\hline
											(8, 8, 16)  & $0.0018$ & $5.8\times 10^{-4}$ & $0.55$ \\
											\hline
											(16, 16, 16) & 0.0166 & 0.0025 & 2.52 \\
											\hline
											(16, 16, 24) &     0.0411 & 0.0025 & Memory Limited \\
											\hline
										\end{tabular}
									\end{table}

									In Table II, we compare the computational time of the proposed DFP receiver, the SIC receiver, and the Projection-SDR algorithm under various system configurations, denoted by $(M_r, K, L)$. The results show that the conventional Projection-SDR method incurs prohibitively high computational overhead, scales poorly, and eventually encounters memory limitations when the block length increases to $L=24$. Compared to the low-complexity SIC receiver, the proposed DFP receiver requires slightly more processing time across the evaluated configurations; for instance, for the system size $(16, 16, 16)$, the DFP receiver takes $0.0166$ seconds, while the SIC receiver requires $0.0025$ seconds. However, this modest increase in computational complexity is justified because the DFP receiver achieves significantly superior detection and estimation performance compared to the SIC baseline. By maintaining an execution time that is orders of magnitude lower than that of SDR while overcoming the performance bottlenecks of SIC, the proposed DFP receiver strikes an attractive balance.

								In Fig. 11, we evaluate the impact of increasing the number of antennas under different system setups. We assume $P_{\rm{r}} = 1$ W and consider a critically determined communication channel, i.e., $K = M_r$. The results show that the number of snapshots significantly affects the detection performance. When the number of snapshots is fixed at $L = 16$, increasing the number of antennas provides only limited performance gain. However, when the number of snapshots increases with the number of antennas, i.e., $L = M_t + 8$, the BER decreases rapidly as the number of antennas increases, following a trend similar to that observed in MIMO communication systems \cite{9018199}. When $K = M_r = 40$ and $\sigma^2 = -20$ dB, the BER approaches $10^{-7}$, which is much lower than that achieved with a fixed number of snapshots. This suggests that the number of snapshots should scale with the number of antennas to fully harness the potential of uplink ISAC systems. Furthermore, as indicated in Lemma 1, increasing the number of transmit antennas used for sensing also increases the BER, further illustrating the inherent tradeoff between S\&C performance.

								\begin{figure*}
																					
									\centering
									\begin{minipage}[t]{0.49\textwidth}
										\centering
										\includegraphics[width=8cm]{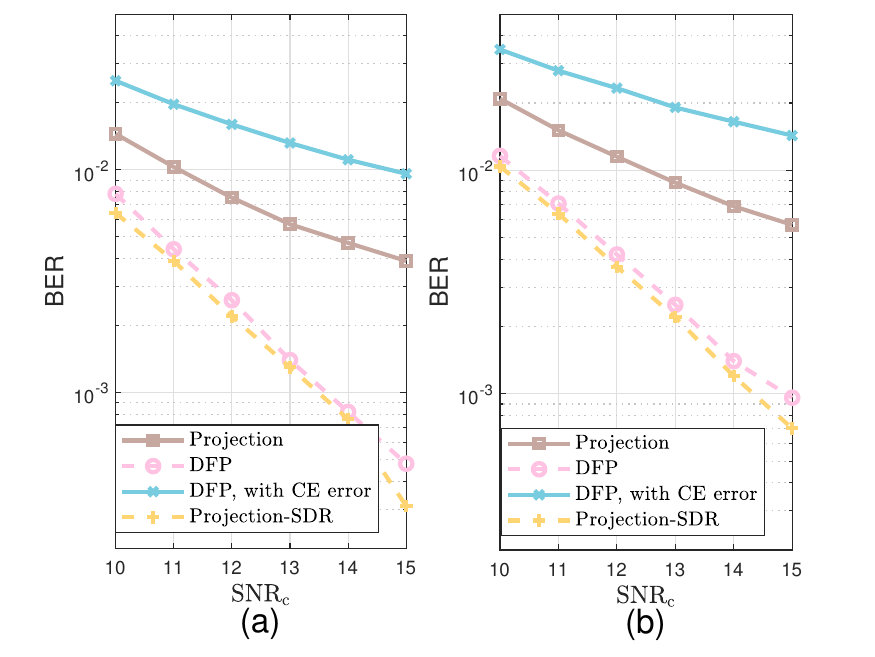}
										\caption*{\small{Fig. 10: BER of different types
										of receivers versus communication SNR under Gaussian channel (a) and correlated channel (b)}}
									\end{minipage}
									\begin{minipage}[t]{0.49\textwidth}
										\centering
										\includegraphics[width=8cm]{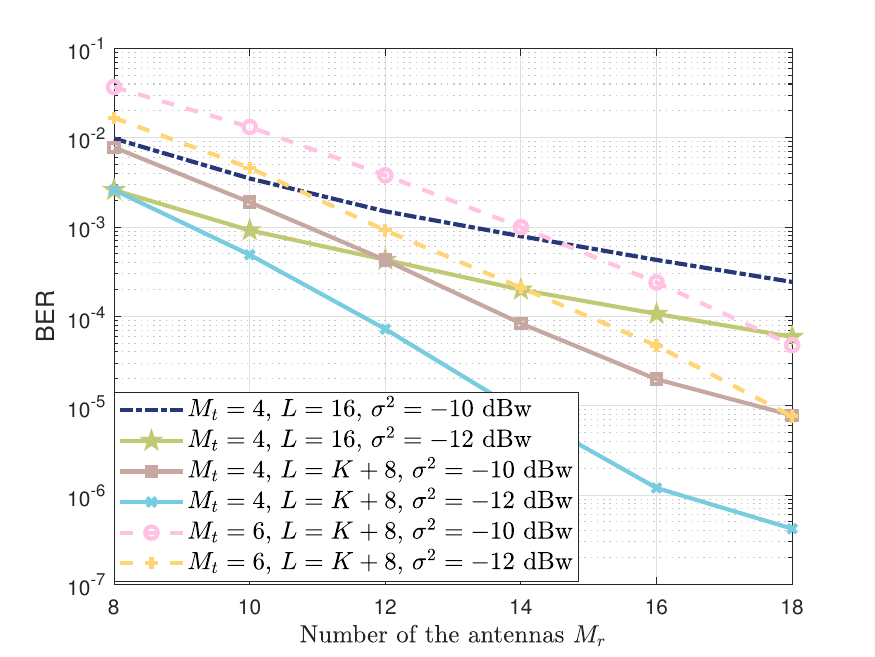}
										\caption*{\small{Fig. 11: BER under different system setups}}
									\end{minipage}
								\end{figure*}

								\section{Conclusion}
								In this paper, the joint signal detection and target response estimation problem in uplink ISAC systems was investigated, and a general receiver framework termed the FP-type receiver with a tradeoff factor was introduced. In the FP-type receiver, communication signals were detected using a reconstructed signal. It was shown that decreasing the tradeoff factor improved the SINR of the signal detection problem but also amplified the undesired correlation between the reconstructed signal. The PEP of the FP-type receiver under both ML and ZF detectors was studied, revealing that the optimal tradeoff factor depends not only on the communication signal detection algorithm but also on the environment. Building on a MIMO signal detection algorithm using homotopy optimization, the DFP-type receiver was proposed, transforming the signal detection problem into a continuous optimization problem. By adjusting the tradeoff factor in each iteration, the DFP-type receiver provided smoother approximations to the original problem and offered improved environmental adaptability. Finally, simulation results verified the necessity of jointly processing an increased number of snapshots.
								
								\numberwithin{equation}{section}
								
								\begin{appendices}

							\section {Proof for Lemma 2}
							The SINR of the transformed signal detection problem using the FP-type receiver is given by 
							\begin{equation} \label{vdfbb}
								\begin{aligned}
									\textrm{SINR}_{\rm{FP}}&\triangleq \frac{\mathbb{E}\left[ {\|\mathbf{\Gamma}_{\rm{FP}} \mathbf{A}_{\rm{c}}\mathbf{x}_{\rm{c}}\|_2^2}\right] }{\mathbb{E}\left[ {\|\mathbf{\Gamma}_{\rm{FP}} \mathbf{A}_{\rm{r}}\mathbf{h}_{\rm{r}}\|_2^2}\right] + \mathbb{E}\left[  \|\mathbf{\Gamma}_{\rm{FP}}\mathbf{n}\|_2^2\right]} \\
									&\overset{(a)}{=} \frac{\mathbb{E}\left[ {\|\mathbf{\Gamma}_{\rm{FP}}\mathbf{A}_{\rm{c}}\mathbf{x}_{\rm{c}}\|_2^2}\right] }{\rho^2 \mathbb{E}\left[ {\| \mathbf{A}_{\rm{r}}\mathbf{h}_{\rm{r}}\|_2^2}\right]+ \mathbb{E}\left[ {\|\mathbf{\Gamma}_{\rm{FP}}\mathbf{n}\|_2^2}\right]} \\
									&= \frac{\mathbb{E}\left[\operatorname{Tr}(\mathbf{\Gamma}_{\rm{FP}}\mathbf{A}_{\rm{c}}\mathbb{E}\left[\mathbf{x}_{\rm{c}}\mathbf{x}_{\rm{c}}^{\rm{H}}\right]\mathbf{A}_{\rm{c}}^{\rm{H}}\mathbf{\Gamma}_{\rm{FP}}^{\rm{H}})\right]}{\rho^2\mathbb{E}\left[ {\| \mathbf{H}_{\rm{r}}\mathbf{X}_{\rm{r}}\|_F^2}\right]+\operatorname{Tr}(\mathbf{\Gamma}_{\rm{FP}}\mathbb{E}\left[\mathbf{n}\mathbf{n}^{\rm{H}}\right]\mathbf{\Gamma}_{\rm{FP}}^{\rm{H}})} \\
									&\overset{(b)}{=}\frac{P_{\rm{c}}\mathbb{E}\left[\operatorname{Tr}(\mathbf{G}_{\rm{FP}} \mathbf{G}_{\rm{FP}}^{\rm{H}})\right]}{\rho^2 L P_{\rm{s}}+ \operatorname{Tr}(\mathbf{\Gamma}_{\rm{FP}} \mathbf{ \Gamma}_{\rm{FP}}^{\rm{H}}) \sigma^2},
								\end{aligned}
							\end{equation}
							where (a) uses the property of the projection matrix and (b) exploits the distribution of the uplink communication signal. Then, we have
							\begin{align} \label{vdfb}
								(\ref{vdfbb})&=\frac{P_{\rm{c}}\mathbb{E}\left[\operatorname{Tr}(\mathbf{G}_{\rm{FP}} \mathbf{G}_{\rm{FP}}^{\rm{H}})\right]}{\rho^2 L P_{\rm{s}}+ \operatorname{Tr}(\mathbf{\Gamma}_{\rm{FP}} \mathbf{ \Gamma}_{\rm{FP}}^{\rm{H}}) \sigma^2} \nonumber \\ 
								&=\frac{P_{\rm{c}}\mathbb{E}\left[ \operatorname{Tr}(\mathbf{P}_{ \rm{FP}} \mathbf{P}_{ \rm{FP}} ^{\rm{H}}\otimes \mathbf{H}_{\rm{c}}\mathbf{H}_{\rm{c}}^{\rm{H}})\right]}{\rho^2 L P_{\rm{s}}+ \operatorname{Tr}(\mathbf{P}_{ \rm{FP}} \mathbf{P}_{ \rm{FP}}^{\rm{H}} \otimes \mathbf{I}_{{M}_r})  \sigma^2} \nonumber \\ 
								&=\frac{P_{\rm{c}}\operatorname{Tr}(\mathbf{P}_{ \rm{FP}} \mathbf{P}_{ \rm{FP}}^{\rm{H}}) \mathbb{E}\left[\operatorname{Tr}(\mathbf{H}_{\rm{c}}\mathbf{H}_{\rm{c}}^{\rm{H}})\right]}{\rho^2 L P_{\rm{s}}+\operatorname{Tr}(\mathbf{P}_{ \rm{FP}} \mathbf{P}_{ \rm{FP}}^{\rm{H}}) \operatorname{Tr}(\mathbf{I}_{{M}_r}) \sigma^2} \nonumber \\ 
								&=\frac{P_{\rm{c}}((1-\rho^2)\operatorname{Tr}(\mathbf{P}_{\perp})+\rho^2 \operatorname{Tr}(\mathbf{I}_L)) K M_r}{\rho^2 L P_{\rm{s}}+((1-\rho^2)\operatorname{Tr}(\mathbf{P}_{\perp})+\rho^2 \operatorname{Tr}(\mathbf{I}_L)){{M}_r} \sigma^2} \nonumber \\ 
								&=\frac{P_{\rm{c}}(L-(1-\rho^2)M_t) K M_r}{\rho^2 L P_{\rm{s}}+(L-(1-\rho^2)M_t)M_r  \sigma^2}.
							\end{align}	
							
							It can be easily observed that $\textrm{SINR}_{\rm{FP}}$ is a monotonically decreasing function with respect to $g(\rho^2)={L\rho^2}/({L-(1-\rho^2)M_t})$, where $\rho^2 \in [0,1]$ and $L \ge M_t$. By taking the first-order partial derivative of $g(\rho^2)$ with respect to $\rho^2$, we have
							\begin{equation}
								\frac{\partial g}{\partial \rho^2} = \frac{L(L - M_t)}{\left(L - (1-\rho^2)M_t\right)^2} > 0.
							\end{equation}
							Therefore, $\textrm{SINR}_{\rm{FP}}$ is a decreasing function with respect to $\rho$. $\hfill\blacksquare$
							
								\section {Proof for Lemma 3 and Lemma 4}
								
								\subsection{Proof of Lemma 3}
								According to the PEP expression for the ML detector, we have
								\begin{equation}
									P_{\rm{ML}}(\tilde{\mathbf{x}}_{\rm{c}} \to \bar{\mathbf{x}}_{\rm{c}})=Q\left( \frac{\|\mathbf{H}_{\rm{c}}\boldsymbol{\Delta }\mathbf{P}_{\rm{FP}}^{*}\|_F}{\sqrt{2\sigma_{\rm{ML}}^2}} \right).
								\end{equation}
								By adopting Assumption 1, i.e., $\mathbf{H}_{\rm{c}}^{\rm{H}}\mathbf{H}_{\rm{c}} \approx \mathbf{I}_{K}$, the Frobenius norm term in the numerator can be approximated as $\|\mathbf{H}_{\rm{c}}\boldsymbol{\Delta }\mathbf{P}_{\rm{FP}}^{*}\|_F \approx \|\boldsymbol{\Delta }\mathbf{P}_{\rm{FP}}^{*}\|_F$. 
								
								According to Assumption 2, the error matrix $\boldsymbol{\Delta }$ contains only a single non-zero element $d_{\rm{min}}$ at the $(p,q)$-th position. Consequently, the matrix product $\boldsymbol{\Delta }\mathbf{P}_{\rm{FP}}^{*}$ results in a matrix where only the $p$-th row is non-zero, and it equals $d_{\rm{min}} [\mathbf{P}_{\rm{FP}}^{*}]_{q, :}$, where $[\mathbf{P}_{\rm{FP}}^{*}]_{q, :}$ denotes the $q$-th row vector of $\mathbf{P}_{\rm{FP}}^{*}$. Thus, we have
								\begin{equation}
									\|\boldsymbol{\Delta }\mathbf{P}_{\rm{FP}}^{*}\|_F = d_{\rm{min}} \left\| [\mathbf{P}_{\rm{FP}}^{*}]_{q, :} \right\|_2.
								\end{equation}
								The conditional PEP for a single-symbol error event occurring at the $q$-th snapshot is then given by
								\begin{equation}
									P_{\rm{ML}}^{(q)} = Q\left( \frac{ d_{\rm{min}} \left\| [\mathbf{P}_{\rm{FP}}^{*}]_{q, :} \right\|_2 }{ \sqrt{2\sigma_{\rm{ML}}^2} } \right).
								\end{equation}
								To evaluate the expected PEP over all possible snapshot indices $q \in \{1, \dots, L\}$, we apply the approximation $\mathbb{E}[Q(x)] \approx Q(\sqrt{\mathbb{E}[x^2]})$ to obtain a closed-form analytical bound. The average squared Euclidean distance over all $L$ snapshots is
								\begin{equation} \label{rms_dist}
									\frac{1}{L} \sum_{q=1}^{L} d_{\rm{min}}^2 \left\| [\mathbf{P}_{\rm{FP}}^{*}]_{q, :} \right\|_2^2 = \frac{d_{\rm{min}}^2}{L} \|\mathbf{P}_{\rm{FP}}^{*}\|_F^2.
								\end{equation}
								Substituting (\ref{rms_dist}) into the Q-function and using the property of the FP-type waveform matrix $\|\mathbf{P}_{\rm{FP}}^{*}\|_F^2 = L - (1-\rho^2)M_t$, we obtain
								\begin{align}
									\mathbb{E}\left[P_{\rm{ML}}(\tilde{\mathbf{x}}_{\rm{c}} \to \bar{\mathbf{x}}_{\rm{c}})\right] &\approx Q\left( \frac{ \sqrt{\frac{d_{\rm{min}}^2}{L} \|\mathbf{P}_{\rm{FP}}^{*}\|_F^2} }{ \sqrt{2\sigma_{\rm{ML}}^2} } \right) \nonumber \\
									&= Q\left( \frac{ d_{\rm{min}}\sqrt{L-(1-\rho^2)M_t}}{ \sqrt{2L \sigma_{\rm{ML}}^2}} \right).
								\end{align}
								This completes the proof of Lemma 3. $\hfill\blacksquare$

								\subsection{Proof of Lemma 4}
	As can be seen from equation (\ref{nfnnf}), the ZF detector completely eliminates the mutual interference between transmitted symbols, thereby decoupling the MIMO channel into $KL$ parallel and independent subchannels. For the $i$-th decoupled subchannel, 	we have $\hat{x}_{{\rm{c}}, i} = x_{{\rm{c}}, i} + w_i$, where $w_i$ denotes the $i$-th entry of the post-processing noise vector $\mathbf{w} = \mathbf{G}_{\rm{FP}}^{\dagger}\hat{\mathbf{n}}$. The covariance matrix of $\mathbf{w}$ is derived as
	\begin{equation}
		\mathbf{R}_{\rm{ZF}} = \mathbb{E}[\mathbf{w}\mathbf{w}^{\rm{H}}] = \sigma_{\rm{ML}}^2 (\mathbf{G}_{\rm{FP}}^{\rm{H}}\mathbf{G}_{\rm{FP}})^{-1}.
	\end{equation}
	By leveraging the Kronecker product structure $\mathbf{G}_{\rm{FP}} = \mathbf{P}_{\rm{FP}}^{\rm{H}} \otimes \mathbf{H}_{\rm{c}}$ and invoking Assumption 1 (i.e., $\mathbf{H}_{\rm{c}}^{\rm{H}}\mathbf{H}_{\rm{c}} \approx \mathbf{I}_K$), the above covariance matrix can be simplified to
	\begin{equation}
		\mathbf{R}_{\rm{ZF}} \approx \sigma_{\rm{ML}}^2 \left[ (\mathbf{P}_{\rm{FP}}\mathbf{P}_{\rm{FP}}^{\rm{H}})^{-1} \otimes \mathbf{I}_K \right].
	\end{equation}
	Accordingly, the equivalent noise variance of the $i$-th subchannel, denoted as $\sigma_{{\rm{ZF}}, i}^2$, corresponds to the $i$-th diagonal entry of $\mathbf{R}_{\rm{ZF}}$, namely $\sigma_{{\rm{ZF}}, i}^2 = [\mathbf{R}_{\rm{ZF}}]_{i,i}$.
	
	The average PEP over all $KL$ symbols is obtained by averaging the individual error probabilities of all decoupled subchannels, which yields
	\begin{equation} \label{exact_pep}
		\bar{P}_{\rm{ZF, exact}} = \frac{1}{K L} \sum_{i=1}^{K L} Q\left( \frac{d_{\rm{min}}}{\sqrt{2 \sigma_{{\rm{ZF}}, i}^2}} \right).
	\end{equation}
	To derive a tractable analytical upper bound, we approximate the exact average PEP by evaluating the Q-function at the average post-processing noise variance across all subchannels, denoted as $\bar{\sigma}_{\rm{ZF}}^2$. This approximation gives rise to
	\begin{equation}
		\mathbb{E}\left[P_{\rm{ZF}}(\tilde{\mathbf{x}}_{\rm{c}} \to \bar{\mathbf{x}}_{\rm{c}})\right] \approx Q\left( \frac{d_{\rm{min}}}{\sqrt{2 \bar{\sigma}_{\rm{ZF}}^2}} \right),
	\end{equation}
	where the average post-processing noise variance $\bar{\sigma}_{\rm{ZF}}^2$ is defined as the normalized trace of the covariance matrix $\mathbf{R}_{\rm{ZF}}$, expressed as
\begin{align} \label{zf_trace}
	\bar{\sigma}_{\rm{ZF}}^2 &= \frac{1}{K L} \sum_{i=1}^{K L} \sigma_{{\rm{ZF}}, i}^2 = \frac{1}{K L} \operatorname{Tr}(\mathbf{R}_{\rm{ZF}}) \nonumber \\
	&\approx \frac{\sigma_{\rm{ML}}^2}{K L} \operatorname{Tr}\left( (\mathbf{P}_{\rm{FP}}\mathbf{P}_{\rm{FP}}^{\rm{H}})^{-1} \otimes \mathbf{I}_K \right) \nonumber \\
	&\stackrel{\text{(a)}}{=} \frac{\sigma_{\rm{ML}}^2}{K L} \operatorname{Tr}\left( (\mathbf{P}_{\rm{FP}}\mathbf{P}_{\rm{FP}}^{\rm{H}})^{-1} \right) \operatorname{Tr}(\mathbf{I}_K) \nonumber \\
	&\stackrel{\text{(b)}}{=} \frac{\sigma_{\rm{ML}}^2}{L} \|\mathbf{P}_{\rm{FP}}^{\dagger}\|_F^2.
\end{align}
	Here, step (a) follows from the trace property of the Kronecker product $\operatorname{Tr}(\mathbf{A} \otimes \mathbf{B}) = \operatorname{Tr}(\mathbf{A})\operatorname{Tr}(\mathbf{B})$, while step (b) holds by substituting $\operatorname{Tr}(\mathbf{I}_K) = K$ and invoking the Frobenius norm definition of the pseudo-inverse.
	
	Substituting the result of (\ref{zf_trace}) into the approximate Q-function expression, and leveraging the inherent property of the FP-type projection matrix $\|\mathbf{P}_{\rm{FP}}^{\dagger}\|_F^2 = L - M_t + \frac{M_t}{\rho^2}$, we arrive at
	\begin{align}
		\mathbb{E}\left[P_{\rm{ZF}}(\tilde{\mathbf{x}}_{\rm{c}} \to \bar{\mathbf{x}}_{\rm{c}})\right] &\approx Q\left( \frac{ d_{\rm{min}} \sqrt{L} }{\sqrt{2\sigma_{\rm{ML}}^2 \|\mathbf{P}_{\rm{FP}}^{\dagger}\|_F^2}} \right) \nonumber \\
		&= Q\left( \frac{ d_{\rm{min}} \sqrt{L} }{ \sqrt{2\sigma_{\rm{ML}}^2 \left(L-M_t+\frac{M_t}{\rho^2}\right)}} \right).
	\end{align}
	This concludes the proof of Lemma 4. \hfill$\blacksquare$

								\section {Proof for Lemma 5}
								Since we have 
								\begin{equation}
									\overline{\bf{G}}^{\rm{T}}(\rho) \overline{\bf{G}}(\rho) =
									\begin{bmatrix}
										{\bf{G}}^{\rm{H}}_{\rm{FP}} {\bf{G}}_{\rm{FP}} & \bf{0} \\
										\bf{0} & {\bf{G}}^{\rm{H}}_{\rm{FP}} {\bf{G}}_{\rm{FP}}
									\end{bmatrix}
								\end{equation}
								and $\lambda_\textrm{1}(\mathbf{P}_{ \rm{FP}}^{\rm{H}}\mathbf{P}_{ \rm{FP}})=1$.
								The largest eigenvalue of matrix $\overline{\bf{G}}^{\rm{T}}(\rho) \overline{\bf{G}}(\rho)$ is given by
								\begin{align}
									\lambda_\textrm{1}(\overline{\bf{G}}^{\rm{T}}(\rho) \overline{\bf{G}}(\rho))&= 	\lambda_\textrm{1}({\bf{G}}^{\rm{H}}_{\rm{FP}} \mathbf{G}_{ \rm{FP}})\\
									&=\lambda_\textrm{1}(\mathbf{P}_{ \rm{FP}}^{\rm{H}}\mathbf{P}_{ \rm{FP}})\lambda_\textrm{1}(\mathbf{H}_{\rm{c}}^{\rm{H}}\mathbf{H}_{\rm{c}})=\lambda_\textrm{1}(\mathbf{H}_{\rm{c}}^{\rm{H}}\mathbf{H}_{\rm{c}}). \nonumber
								\end{align}
								According to Theorem 2, it can be proved that Problem (\ref{cst}) is equivalent to the transformed signal detection problem in (\ref{fbgf}). Finally, when $\rho=0$, the FP-type receiver reduces to the conventional projection-type receiver, which is the same as the problem formulated in (\ref{vbdf}).

							\end{appendices}


							
							
							%
							
								
								
							\bibliographystyle{IEEEtran}
							\bibliography{IEEEabrv,AMI4}

\begin{IEEEbiography}[{\includegraphics[width=1in,height=1.25in,clip,keepaspectratio]{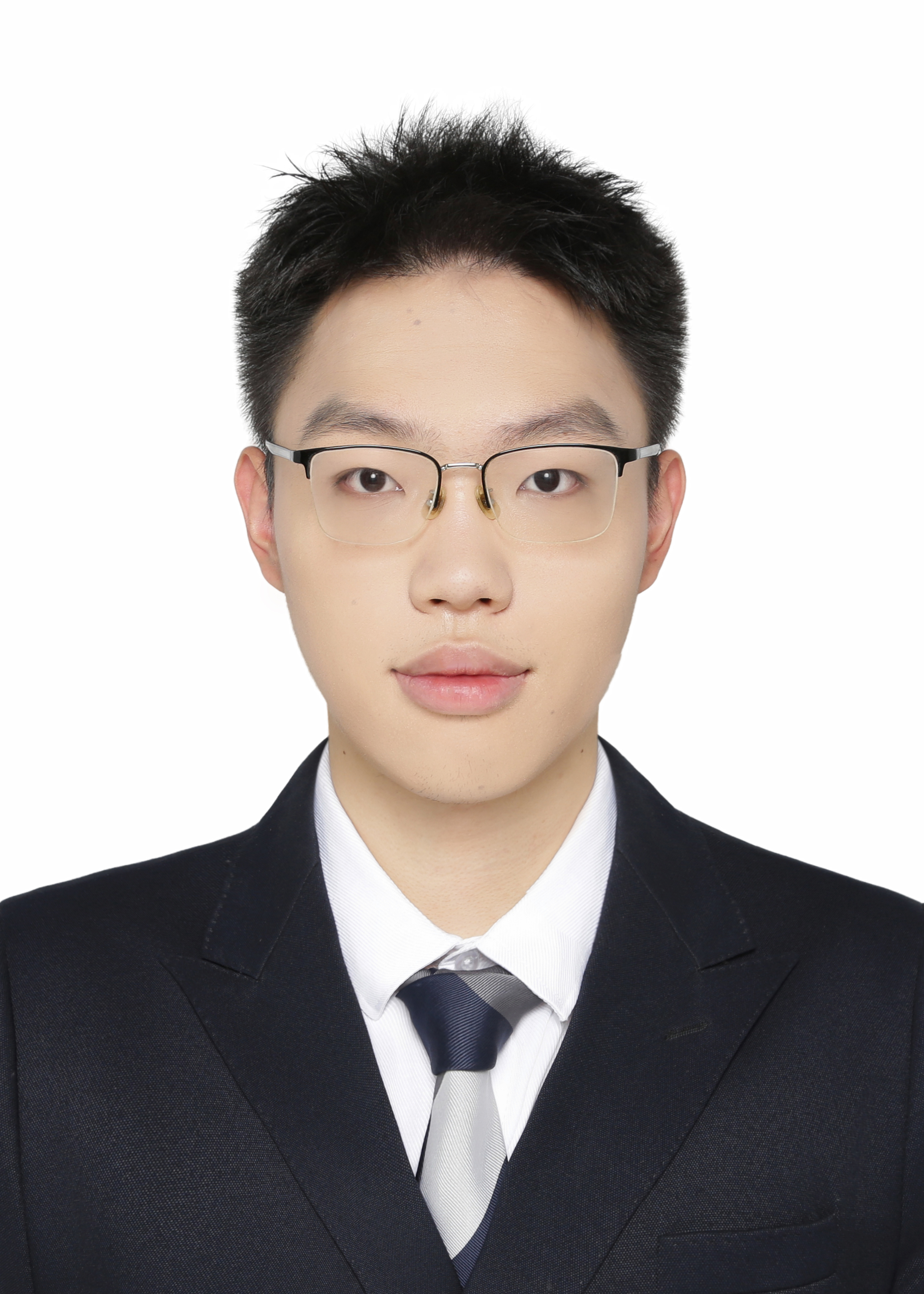}}]{Zhiyuan Yu}
received the B.S. degree from the School of Information Science and Engineering, Southeast University, Nanjing, China, in 2023. He is currently pursuing the M.S. degree with the same institution. He was the recipient of the inaugural President’s Graduate Scholarship of Southeast University in 2024, and the IEEE ComSoc SPCC Best Paper Award in 2025. His research interests include reconfigurable intelligent surfaces (RIS), and integrated sensing and communication (ISAC).
\end{IEEEbiography}

\begin{IEEEbiography}[{\includegraphics[width=1in,height=1.25in,clip,keepaspectratio]{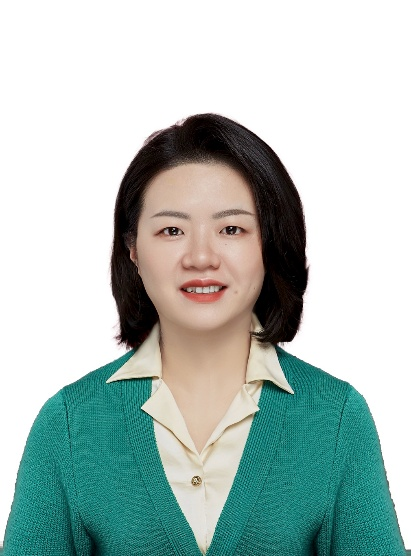}}]{Hong Ren}
	(Member, IEEE) received the B.S. degree from the School of Information Science and Engineering, Southwest Jiaotong University, Chengdu, China, in 2011, and the M.S. and Ph.D. degrees from Southeast University, Nanjing, China, in 2014 and 2018, respectively. From 2016 to 2018, she was a Visiting Student with the University of Southampton, U.K., and from 2018 to 2020, a Post-Doctoral Scholar with the Queen Mary University of London, U.K. She is currently an Associate Professor with Southeast University. Her research interests include communication and signal processing, cooperative ISAC, artificial intelligence (AI), and ultra-reliable low-latency communications (URLLC). She was recognized as a Clarivate Highly Cited Researcher in 2024 and 2025. She was a recipient of the IEEE ComSoc Heinrich Hertz Award in 2025, the IEEE ComSoc SPCC Best Paper Award in 2025, the IEEE Communications Society Fred W. Ellersick Prize in 2024, and the IEEE Communications Society Leonard G. Abraham Prize in 2022. She is also an Editor of \textsc{IEEE Transactions on Vehicular Technology} and \textsc{IEEE Wireless Communications Letters}.
\end{IEEEbiography}

\begin{IEEEbiography}[{\includegraphics[width=1in,height=1.25in,clip,keepaspectratio]{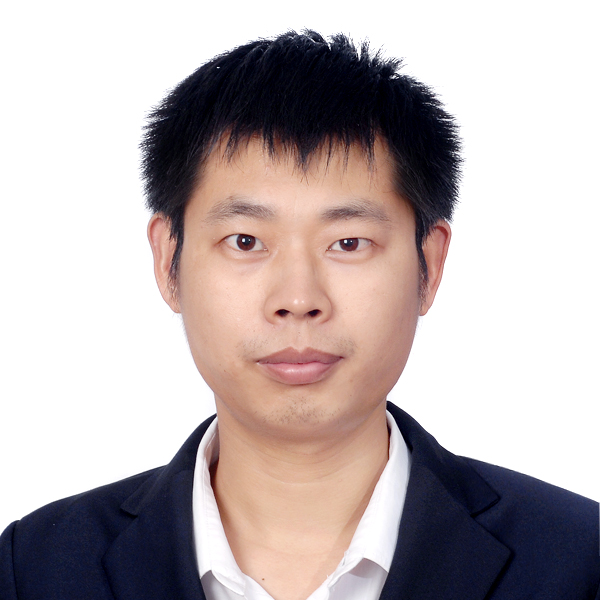}}]{Cunhua Pan}
	(Senior Member, IEEE) is currently a Full Professor with Southeast University. He has published over 200 IEEE journal articles. His papers have received over 24,000 Google Scholar citations with an h-index of 76. His research interests mainly include reconfigurable intelligent surfaces (RIS), AI for wireless, near-field communications and sensing, and integrated sensing and communications. He received the IEEE ComSoc Leonard G. Abraham Prize in 2022, the IEEE ComSoc Asia-Pacific Outstanding Young Researcher Award in 2022, the IEEE ComSoc Fred W. Ellersick Prize in 2024, the IEEE ComSoc CTTC Early Achievement Award in 2024, the IEEE ComSoc SPCC Early Achievement Award in 2024, the IEEE ComSoc RCC Early Achievement Award in 2025, the IEEE ComSoc Heinrich Hertz Award in 2025, and the IEEE ComSoc SPCC Best Paper Award in 2025. One Ph.D. thesis he supervised won the IEEE Signal Processing Society Best Ph.D. Dissertation Award in 2024. He is a Clarivate Highly Cited Researcher. He is/was an Editor of \textsc{IEEE Transactions on Communications}, \textsc{IEEE Transactions on Vehicular Technology}, \textsc{IEEE Wireless Communications Letters}, and \textsc{IEEE Communications Letters}. He serves as the Leading Guest Editor for \textsc{IEEE Journal on Selected Areas in Communications} and \textsc{IEEE Journal of Selected Topics in Signal Processing}.
\end{IEEEbiography}

\begin{IEEEbiography}[{\includegraphics[width=1in,height=1.25in,clip,keepaspectratio]{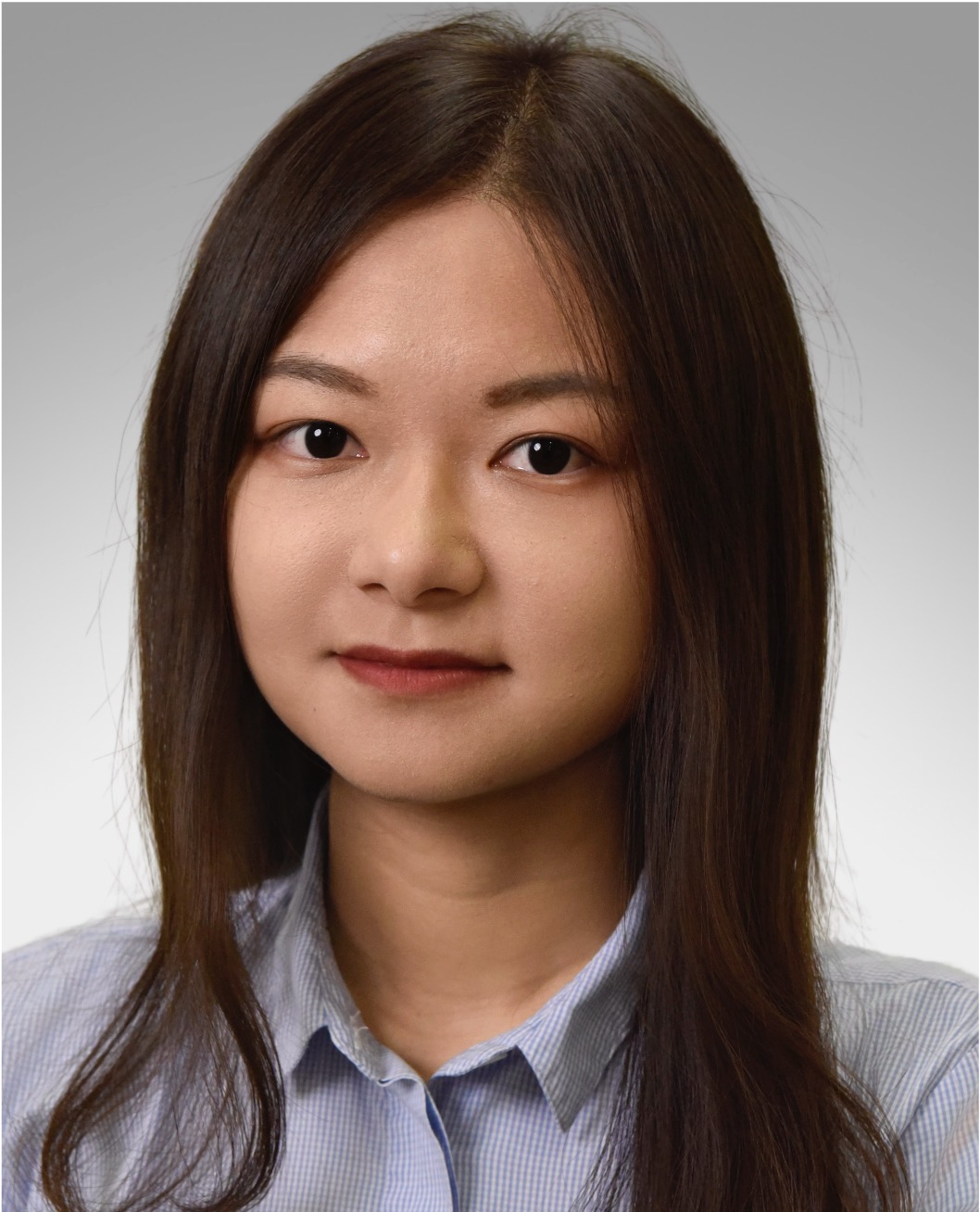}}]{Gui Zhou}
	(Member, IEEE) received the B.S. and M.E. degrees in information and electronics from Beijing Institute of Technology, Beijing, China, in 2015 and 2019, respectively, and the Ph.D. degree in electronic engineering and computer science from Queen Mary University of London, U.K., in 2022. She was a Humboldt Post-Doctoral Research Fellow and a Post-Doctoral Research Fellow with the Institute for Digital Communications, Friedrich-Alexander Universit\"{a}t of Erlangen-N\"{u}remberg (FAU), Erlangen, Germany, from 2022--2025. She is currently a Professor with the School of Electronic Information and Communication, Huazhong University of Science and Technology (HUST), Wuhan, China. Her major research interests include channel estimation, transceiver design, integrated sensing and communication, and array signal processing. She received the 2025 Best Paper Award from the IEEE ComSoc SPCC, the 2024 IEEE SPS Best Ph.D. Dissertation Award, and the Best Paper Award for WCSP2022. She is an Editor of \textsc{IEEE Transactions on Communications}.
\end{IEEEbiography}

\begin{IEEEbiography}[{\includegraphics[width=1in,height=1.25in,clip,keepaspectratio]{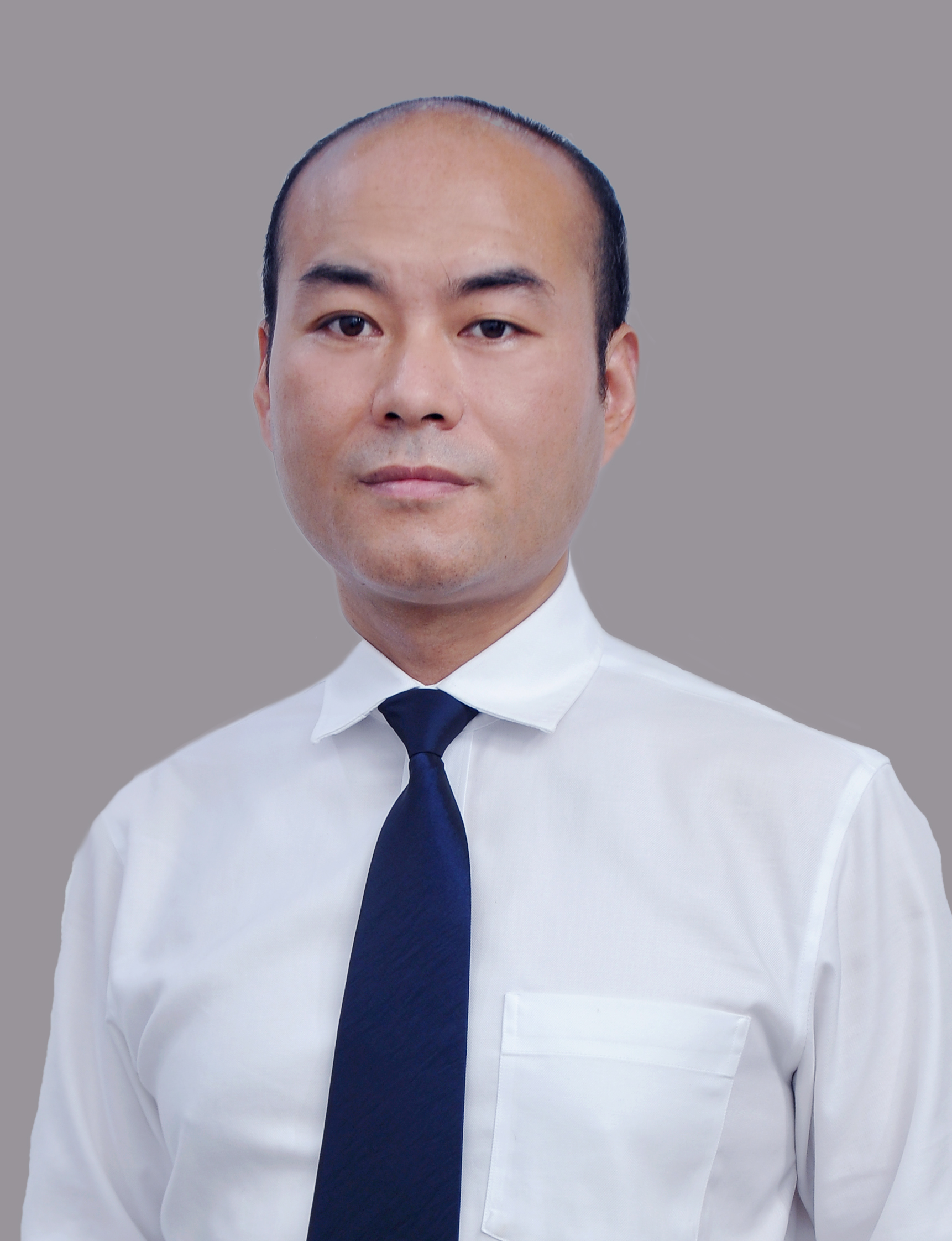}}]{Dongming Wang}
	(Member, IEEE) received the B.S. degree from Chongqing University of Posts and Telecommunications in 1999, the M.S. degree from the Nanjing University of Posts and Telecommunications in 2002, and the Ph.D. degree from Southeast University, China, in 2006. In 2006, he joined the National Mobile Communications Research Laboratory, Southeast University, where he is currently a Professor. His research interests include signal processing for wireless communications and large-scale distributed MIMO systems (cell-free massive MIMO). He served as the Symposium Co-Chair for the 2015 IEEE International Conference on Communications (ICC 2015) and the IEEE Wireless Communications and Signal Processing Conference (IEEE WCSP 2017).
\end{IEEEbiography}

\begin{IEEEbiography}[{\includegraphics[width=1in,height=1.25in,clip,keepaspectratio]{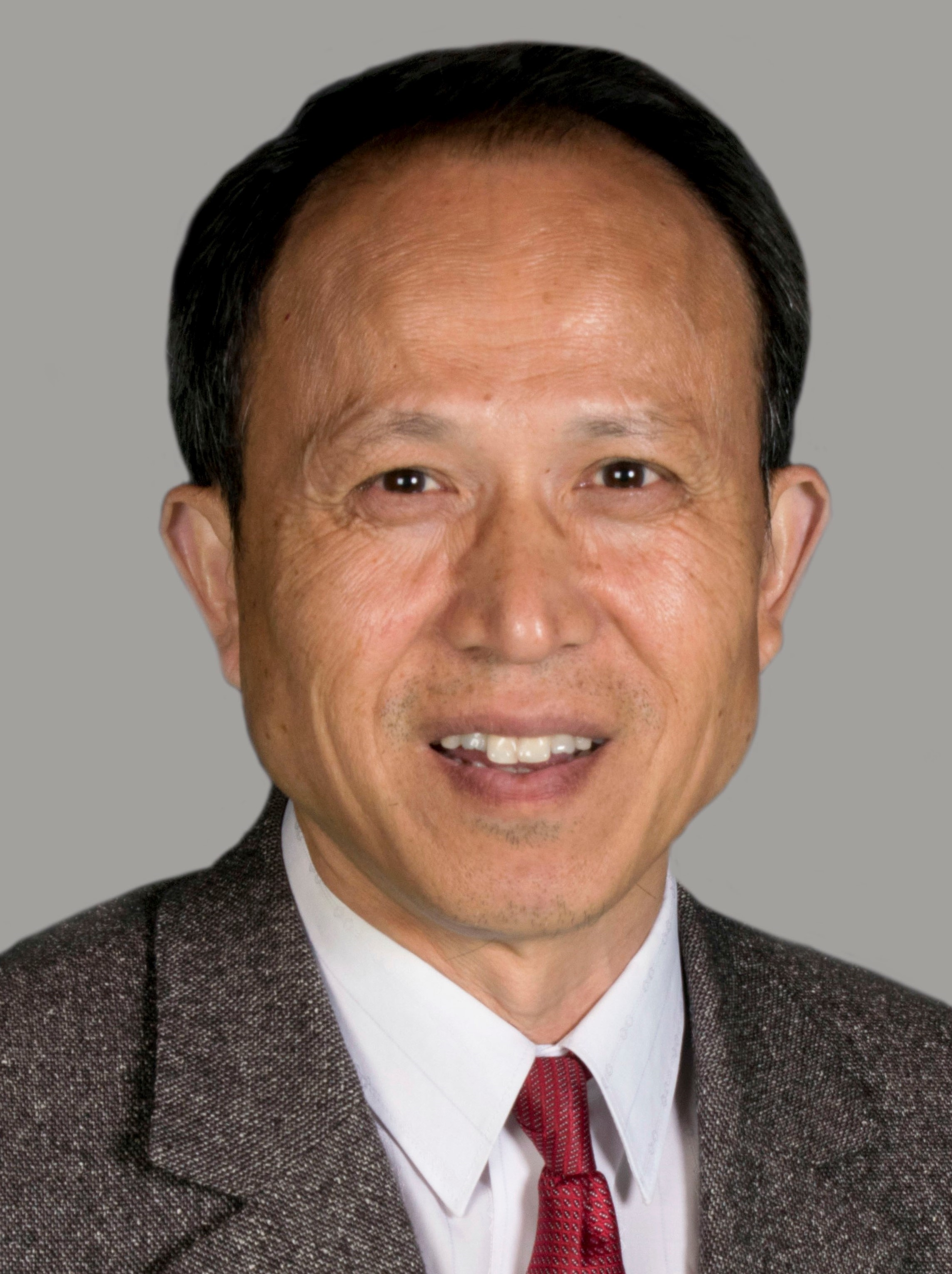}}]{Jiangzhou Wang}
	(Fellow, IEEE) is currently a Professor with Southeast University and an Emeritus Professor with the University of Kent. He has published more than 600 articles and five books. His research interests include mobile communications. He is an International Member of Chinese Academy of Engineering (CAE) and a fellow of the Royal Academy of Engineering (RAEng), U.K. He was a recipient of the 2024 IEEE Communications Society Fred W. Ellersick Prize and the 2022 IEEE Communications Society Leonard G. Abraham Prize. He is the General Chair of the 2026 IEEE Global Communications Conference (GLOBECOM2026), Macau. He was the Technical Program Chair of the 2019 IEEE International Conference on Communications (ICC2019), Shanghai, the Executive Chair of the IEEE ICC2015, London, and the Technical Program Chair of the IEEE WCNC2013.
\end{IEEEbiography}

						\end{document}